# Restoring self-limited growth of single-layer graphene on copper foil via backside coating

Nicolas Reckinger,[1,*] Marcello Casa,[2] Jeroen E. Scheerder,[3] Wout Keijers,[3] Matthieu Paillet,[4] Jean-Roch Huntzinger,[4] Emile Haye,[1] Alexandre Felten,[5] Joris Van de Vondel,[3] Maria Sarno,[2] Luc Henrard,[1] and Jean-François Colomer[1]

[1] Department of Physics, University of Namur, Rue de Bruxelles 61, 5000 Namur, Belgium.

[2] Department of Industrial Engineering and Centre NANO_MATES, University of Salerno Narrando Srl, Via Giovanni Paolo II, 132, 84084 Fisciano, Italy.

[3] Department of Physics and Astronomy, Celestijnenlaan 200D, KU Leuven, Leuven B-3001, Belgium.

[4] Laboratoire Charles Coulomb (L2C), Université de Montpellier, CNRS, Montpellier, France.

[5] SIAM platform, University of Namur, Rue de Bruxelles 61, 5000 Namur, Belgium.

Email: nicolas.reckinger@gmail.com

**Abstract**

The growth of single-layer graphene (SLG) by chemical vapor deposition (CVD) on copper surfaces is very popular because of the self-limiting effect that prevents the growth of few-layer graphene (FLG). However, the reproducibility of the CVD growth of homogeneous SLG remains a major challenge, especially if one wants to avoid heavy surface treatments, monocrystalline substrates and expensive equipment to control the atmosphere inside the growth system. We demonstrate here that backside tungsten coating of copper foil allows the exclusive growth of SLG with full coverage by atmospheric pressure CVD implemented in a vacuum-free outfit. We show that the absence of FLG patches is related to the absence of decomposition of methane on the backside and consequently to the suppression of C diffusion through copper. In the perspective of large-scale production of graphene, this approach constitutes a significant improvement to the traditional CVD growth process since (1) a tight control of the hydrocarbon flow is no longer required to avoid FLG formation and, consequently, (2) the growth duration necessary to reach full coverage can be dramatically shortened.

Chemical vapor deposition (CVD) has become the most popular production method for graphene, mainly because it holds great promises for industrial-scale applications. Catalytic CVD is a conceptually simple technique: it involves the decomposition of hydrocarbon precursors on substrates at high temperature in a controlled atmosphere at low[1] or atmospheric pressure.[2] In particular, copper (Cu) is extensively chosen as a substrate because it allows self-limited graphene growth due to its very low carbon (C) solubility, leading to highly homogeneous graphene sheets.[1]

The main focus of the recent research devoted to CVD growth of graphene is to produce ever larger graphene single crystals aiming, notably, to eliminate the detrimental effect of graphene domain boundaries on electron transport. The dominant approach towards this goal is to decrease the nucleation density of graphene by suppressing or passivating the nucleation sites (defects and surface steps at Cu's surface, impurities, etc.) by various treatments: chemical mechanical polishing;[3] electropolishing (EP);[4] prolonged thermal annealing;[5] high-pressure thermal annealing;[6] melting and resolidification;[7] pre-growth superficial oxidation;[8,9,10,11,12,13,14,15,16] surface engineering with melamine;[17] oxygen-assisted growth;[18] second passivation[19] and oxygen-assisted C contamination scavenging.[20] A second, less popular technique is to grow the graphene flakes in epitaxial registry with a monocrystalline Cu substrate. In consequence, the domains are aligned relatively to each other and merge seamlessly to produce graphene sheets free of domain boundaries. Such monocrystalline substrates can be obtained from the epitaxial deposition of thin Cu films on various kinds of single crystals.[21,22,23,24,25] However, it is more convenient and cost-effective to start from cold-rolled polycrystalline Cu foils and convert them (at the surface or in the bulk) into monocrystals by appropriate strategies such as a prolonged thermal annealing at high temperature,[26,27] successive oxidative and reductive annealing at high temperature,[28,29] the hole-pocket method[30], or a Czochralski-like reconstruction induced through a temperature gradient.[31] A completely different route consists of working with the smooth surface of melted Cu.[32]

A very important challenge is related to the unwanted formation of few-layer graphene (FLG) domains inside the large-sized single-layer graphene (SLG) flakes or films. Even though SLG CVD growth on Cu is in principle self-limited, the presence of impurities or defects acting as nucleation centers breaks down this behavior[3], more specifically in atmospheric pressure conditions. The C-rich molecules that decompose on the frontside Cu surface are often regarded as the source for the FLG nucleation.[33] FLG flakes are considered to grow either on top of the first graphene layer, via layer-by-layer epitaxy,[34,35] or underneath, by C intercalation under the first-grown graphene flakes.[36,37] In that respect, C diffusion through the Cu foil is often disregarded as a supplier of carbonaceous species. However, Fang et al.[38] show that, despite C's low solubility in Cu, it can decompose on one face of a Cu enclosure, dissolve in and diffuse through the foil to form FLG flakes under graphene grown on the opposite side. Later, the same group claim that a tungsten (W) foil inserted inside the Cu enclosure can be used as a C sink to inhibit FLG growth.[39] By growing a thin Cu oxide layer on the backside of Cu foils prior to graphene growth, Braeuninger-Weimer et al.[20] also demonstrate how oxygen (O) can diffuse through the Cu foil to scavenge C impurities, thereby enabling a drastic decrease of the graphene nucleation density on the front surface. Recent publications also evidence the complete suppression of FLG patches when a nickel (Ni) substrate (foil or foam) is placed between the fused silica carrier and the flat Cu foil.[40,41] In both cases, Ni acts as a C "getterer" and prevents C diffusion. Finally, Yoo *et al.*[42] deposit a thin layer of Ni on the Cu foil's backside. They find out that the graphene layer number grown on the frontside depends on the thickness of the Ni thin film.

In this work, we propose backside W deposition to obtain CVD growth of strictly homogeneous SLG films under atmospheric pressure with vacuum-free equipment. We focus on reproducibility, which is a major concern in the CVD graphene community, especially when the process is carried out outside the controlled environment of cleanrooms. First, we demonstrate that a range of conventional surface treatments used to clean the Cu foils prior to growth are inefficient to achieve reproducibility in terms of graphene coverage. We also find that these treatments, including EP, are insufficient to suppress the formation of FLG graphene islands. The main novelty of this study is the deposition of a thin W layer on the backside of electropolished Cu foils. We show that this W backside coating leads to the reproducible growth of exclusively SLG. This remarkable result is explained by the complete suppression of C diffusion through the Cu foil, which restores self-limited growth of SLG. The W back coating enables to relax the strict control on the growth conditions, greatly facilitating the production of exclusively SLG sheets at an industrial scale.

## Results and discussion

We have first investigated the reproducibility of SLG CVD growth on Cu foils in terms of coverage and homogeneity under the standard cleaning (see Methods) and growth conditions (see the supporting information (SI), Figures S1−4). Only 50% of the samples (on a total of 25) show complete graphene surface coverage, with a highly variable concentration of FLG inclusions. Based on indirect evidence, we identify two causes for this erratic behavior: (1) residual impurities due to an insufficient surface cleaning, and (2) Cu nanoparticles created during the pre-growth annealing, due to the mild oxidation of the Cu foil's surface under argon (and residual oxidizing impurities) followed by reduction under argon (Ar) and hydrogen ($H_2$) (see Figure S6). The impurities and the Cu particles are both responsible for the FLG nucleation and the partial graphene coverage. In addition, the inherently stochastic nature of the nucleation process complicates a systematic investigation to determine why FLG formation occurs at a specific particle site on the Cu foil and not at others.

In an attempt to reduce the amount of impurities on the Cu foils, several alternative cleaning techniques proposed in the literature are tested. On the one hand, we find that no chemical treatment leads to reproducibility in our conditions (see Figure S5a−d). On the other hand, EP is an appealing preparation technique since it implies the removal of Cu atoms from the surface of the foil together with the contaminants. We show that a clean Cu surface, obtained with EP, gives a reproducible full graphene coverage (see Figure S7c). However, uniformity remains an issue (see Figure S7d), probably due to the formation of Cu nanoparticles during the pre-growth annealing.

Our preliminary results plainly emphasize that none of the chemical and electrochemical treatments we have tried result in complete reproducibility. We then attempt to hinder the FLG nucleation by using a Ni foil, serving as a C sink, to support the Cu substrates.[40,41] As seen in Figure S5e, this trial has failed, possibly because the Cu and Ni foils are not perfectly flat and in intimate contact with each other, leaving gaps for $CH_4$ to diffuse through Cu. Another group has used a W foil inside a Cu enclosure for the same purpose.[39]

To prevent any $CH_4$ diffusion through Cu, we deposit a thin (50 nm) layer of W on the backside of electropolished Cu pieces. The deposition ensures an intimate contact with the Cu foil's backside and the Cu samples remain flat, facilitating the manipulation (as opposed to the Cu enclosure configuration). Since W has a very high melting point (as high as 3422 °C at ambient pressure), the thin film is able to sustain the harsh thermal conditions in the reactor and retain its integrity during the full procedure.

In Figure 1a, a low-magnification SEM image of the surface of a Cu piece after graphene growth (standard growth conditions) with W covering half of the backside (this half is called the "W half" in the following) is shown. Spectacularly enough, the two halves of the Cu foil exhibit very distinct morphological aspects. The left side (W half) of Figure 1a is very heterogeneously contrasted, meaning that Cu remains polycrystalline. The other half reveals a homogeneous morphology, indicative of a Cu reconstruction in the (111) crystalline orientation.[28] The reconstruction of cold-rolled Cu foils occurs at high temperature, via the mechanism of abnormal grain growth, if the grain boundaries are left free to evolve.[28] In the present case, the W layer pins the Cu grain boundaries and prevents any further reconstruction even at temperatures as high as 1050 °C.

It is also worth noting that the graphene flakes on the W-free half are hexagonal, while they exhibit an irregular shape on the other half, probably related to the polycrystallinity of Cu. More importantly, on the W-free half, FLG islands are clearly visible. In contrast, it is difficult to assess the presence of FLG flakes on the W half due to the mosaic of Cu grain orientations. In addition, the density in SLG graphene flakes is significantly lower on the W half, suggesting that W plays a crucial role in the supply of C building blocks to the front surface.

To increase the graphene coverage on the W half and to better visualize the FLG flakes, we produce a similar sample with a slightly higher dilute $CH_4$ flow (0.6 sccm) and we transfer it on a 90-nm-thick silicon dioxide/silicon ($SiO_2$/Si) sample (see Figure 1b). Spectacularly, the W half is completely devoid of FLG graphene islands, while the other half is completely scattered with them. To quantitatively confirm this observation, simultaneous micro-Raman and micro-reflection analyses are performed on the whole sample. Figure 1c displays the number of layers $N_G$ deduced from the normalized Raman G-band area (see the SI). Further characterization and analysis of this sample are available in the SI (see Figures S8−10). It gives a definitive proof that the W part of the sample comprises almost uniquely SLG while the W-free half is very heterogeneous.

Consequently, in our next experiments, since the W layer pins the Cu grain boundaries, we first reconstruct the Cu foil in the (111) orientation after EP, then deposit W on the backside, and finally, grow graphene on the front side of the Cu foil. Figure 1d displays a photograph of four such samples after growth and heating on a hot plate in air at 150 °C for 5 min, corresponding to a progressive increase of the dilute CH$_4$ flow (from 0.6 to 1.2 sccm by steps of 0.2 sccm). We can clearly see that the graphene coverage increases progressively, until it is complete for 1.2 sccm (the reddish color indicates oxidized Cu).

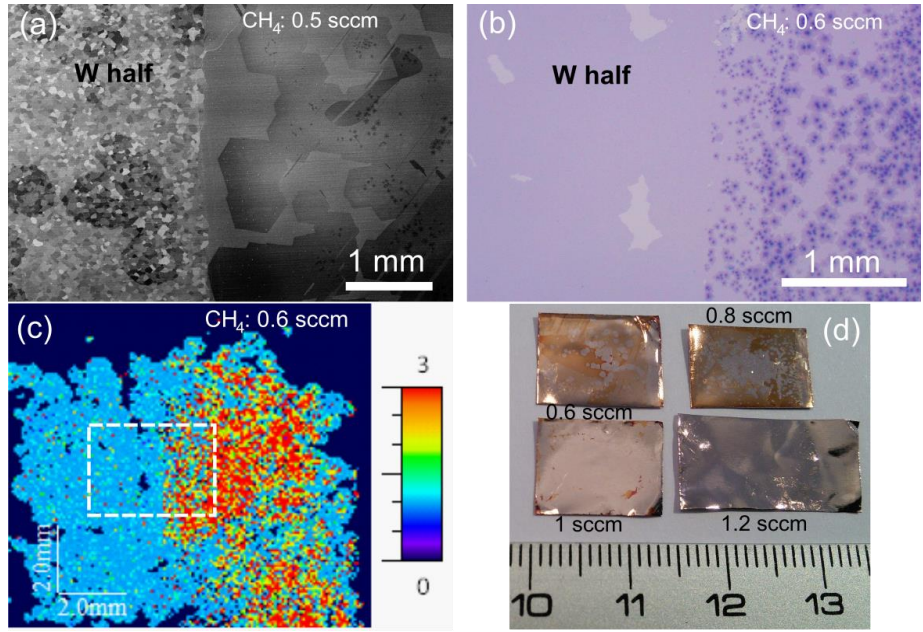

Figure 1: (a) Low-magnification scanning electron microscopy image of a Cu piece after graphene growth with a 50-nm-thick W layer covering the left half of its backside (grown with a 0.5 sccm dilute CH$_4$ flow). (b) Optical microscopy picture of a graphene sample (grown with a 0.6 sccm dilute CH$_4$ flow) transferred onto a 90-nm-thick silicon dioxide/silicon substrate. (c) Map of the number of layers of the sample shown in (b) deduced from the normalized integrated Raman G-band intensity. The dashed white rectangle corresponds to the region shown in (b). (d) Four Cu samples corresponding to a progressive increase of the dilute CH$_4$ flow (from 0.6 to 1.2 sccm by steps of 0.2 sccm), photographed after graphene growth and heating on a hot plate in air to reveal the oxidized, uncovered (reddish) Cu surface.

Next, we perform simultaneous micro-reflection and micro-Raman mapping on the 1.2 sccm sample of Figure 1d transferred onto a 90-nm-thick SiO$_2$/Si substrate (see the SI, Figures S11−13) to give a quantitative support to our claim of an exclusively SLG film. We follow the approach detailed in Ref. [43], with $N_G$ the number of layers obtained using the normalized Raman G-band area, and $N_{OC}$ the number of layers obtained from the laser optical contrast (see Figures 2a and 2b). Both data sets agree and confirm that except for two edges, the sample is almost exclusively composed of SLG. Furthermore, in Figure 2c, the 3-dimensional bivariate histograms of $N_G$ and $N_{OC}$, as well as the histograms for each independent quantity, are displayed for the central part of the sample delimited by the yellow dotted line in Figure 2a. Quantitatively, on the 6500 points where the number of layers has been attributed, 3.1% correspond to the bare substrate (i.e. no graphene), 1.1% are 0−1 layer (only partial graphene coverage), 94.2% are SLG, 1.3% are between 1 and 2 layers (graphene wrinkles, partial bilayer graphene coverage, etc.) and 0.3% is bilayer (mostly small graphene pieces scratched and folded during the transfer).

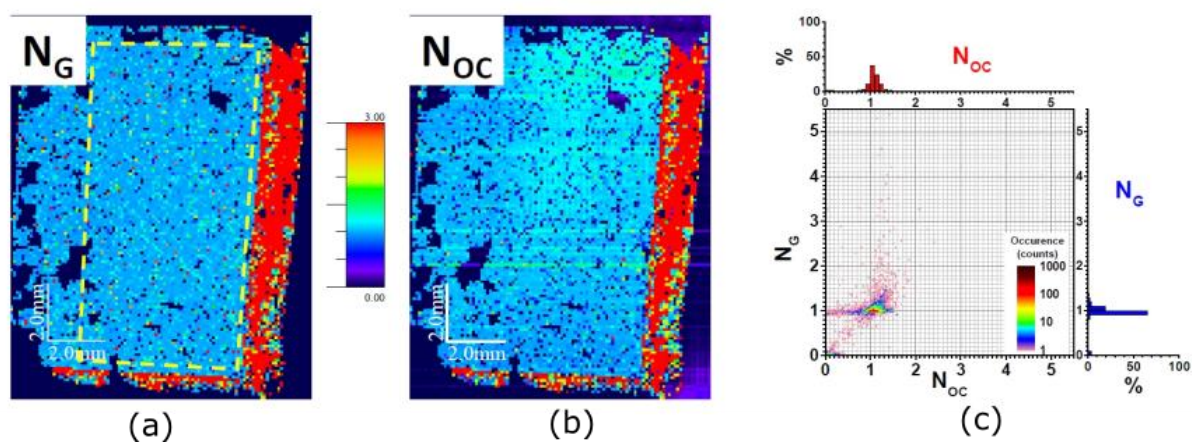

Figure 2: Maps of the number of layers of the sample grown with a 1.2 sccm dilute $CH_4$ flow and transferred on a 90-nm-thick silicon dioxide/silicon substrate (a) $N_G$, the number of layers deduced from the normalized integrated Raman G-band intensity and (b) $N_{OC}$, the number of layers deduced from the laser optical contrast. (c) 3D bivariate histogram (0.025 bin size) of $N_{OC}$ and $N_G$ derived from the maps (a) and (b). The region considered is delimited by the dashed yellow line in (a). The number of occurrences (frequency counts) is color-coded as shown on the graphs. On top (resp. right hand side) are displayed the corresponding histograms of $N_{OC}$ (resp. $N_G$).

We also evaluate the W backside coating method when the $CH_4$ flow is significantly increased (3 sccm, maximal value of our mass flow), with the objective of drastically decreasing the growth duration (fixed here to 5 min, instead of 1 h as before). The corresponding data are given in the SI, Figures S14−16. The Raman spectroscopy/optical contrast mapping results show that the corresponding graphene is also exclusively monolayer, with an almost complete coverage, evidencing the robustness of the synthesis technique. This is a very important aspect of the W backside coating approach in the perspective of industrial production since a tight control of the $CH_4$ flow is no longer required to avoid FLG formation and the full-coverage synthesis process duration can be dramatically shortened.

To better understand the precise role of the W coating, we inspect, by X-ray diffraction (Figure 3a) and depth profile XPS (Figure 3b), the chemical and morphological evolution of a W film just after deposition (no thermal treatment) on Cu and after graphene growth, respectively. From Figure 3a, the as-deposited W film appears amorphous, since no W-related diffraction peak can be observed. The graphene growth process leads to the crystallization of metallic W, in the cubic $I_{m\bar{3}m}$ structure (space group 229), with the occurrence of diffraction peaks located at 40.3, 58.2, 73.2, 87, and 100.6° attributed to the (110), (200), (211), (220) and (310) crystallographic orientations, respectively.[44] W is known for its carbide forming capabilities[45] but we find no trace of diffraction peaks related to W carbide. Furthermore, the XPS analysis shows that the as-deposited W layer is slightly oxidized (not shown) while Cu and W appear intermixed after graphene growth (roughly a 50% Cu/50% W alloy) and no C is contained in that layer (within the detection threshold of XPS of <0.5 at% for the acquisition parameters used during the XPS profile), in agreement with Fang *et al.*[39] Graphene can be grown on W foils[46] but, here, C detected at the very surface of the sample corresponds to organic contamination. Based on these two analyses, it seems that the W layer acts as a C diffusion barrier rather than as a C sink, since no C is trapped in the W-bearing layer.

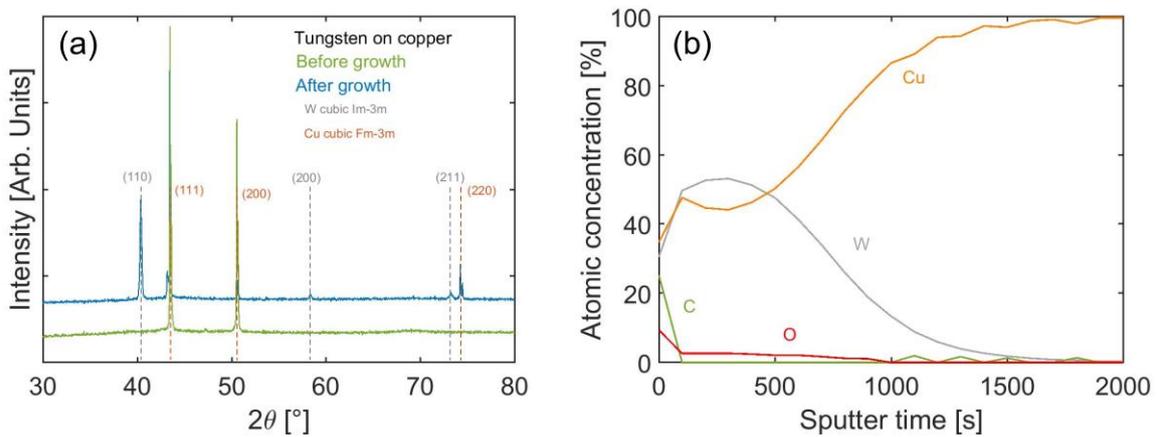

Figure 3: (a) X-ray diffraction analysis of a W film on Cu, just after deposition and after graphene growth. (b) X-ray photoelectron spectroscopy depth profile of a W film on Cu after graphene growth.

Finally, electronic characterization of the as-synthesized graphene (transferred on a 300-nm-thick $SiO_2/p^{++}$ Si substrate) is carried out using a graphene field-effect transistor (GFET) in a Hall bar configuration to evaluate its quality (see Figure S17a). Measurements of the longitudinal resistivity $\rho_{xx} = V_{xx}/I \times W/L$ and transversal (Hall) resistivity $\rho_{xy} = V_{xy}/I$ are conducted with a current bias of 10 µA (see Figure S17b). By sweeping a gate voltage $V_G$, applied between the graphene Hall bar and the $p^{++}$ Si electrode, the Fermi level of graphene can be altered, resulting in the well-known ambipolar field-effect behavior. The sheet conductance $\sigma_{xx} = 1/\rho_{xx}$ versus $V_G$ for a representative GFET is shown in Figure 4a, at room temperature and at 400 mK. In the insets, the electronic level filling of the Dirac cone is schematically indicated by the shaded regions. We find an average electron-hole mobility at both temperatures of ~4 × $10^3$ cm$^2$/(Vs). Hall measurements are performed at 400 mK, at a magnetic field of 5 T applied perpendicular to the graphene plane. The Hall conductivity as a function of $V_G$, shown in Figure 4b, demonstrates the clear half-integer quantum Hall effect, $\sigma_{xy} = 4e^2/h \times (n + 1/2)$ (n = 0, 1, 2,…). This is typical of SLG[46,47] and serves as an indication for the high electronic quality of the sample.[48]

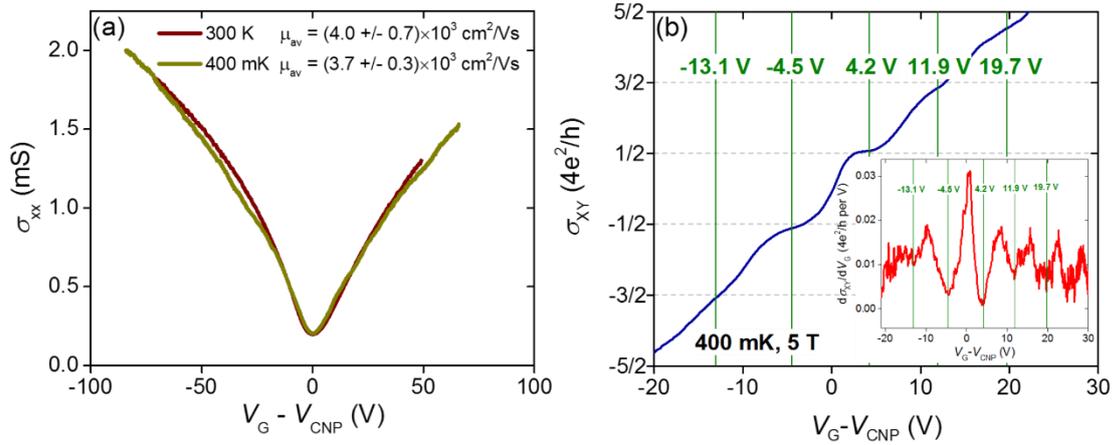

Figure 4: (a) Graphene sheet conductivity as a function of the gate voltage with respect to the charge neutrality point, $\sigma_{xx}(V_G - V_{CNP})$, at room temperature and 400 mK. The insets portray the dispersion bands of graphene, with the shaded areas representing the level filling. (b) Hall conductivity $\sigma_{xy}(V_G - V_{CNP})$ at $B = 5$ T and 400 mK. Plateaus appear at $\sigma_{xy} = 4e^2/h \times (n + 1/2)$, with $n$ an integer. Vertical lines show the gate values, where the first derivative of $\sigma_{xy}$ has local minima (shown in the inset), aiding in the identification of the plateaus.

## Conclusion

The conditions to obtain a uniform, full-coverage SLG sheet grown by CVD on Cu substrates in vacuum-free equipment have been examined in detail. Usual surface cleaning techniques are found to leave residual contaminations and nanoparticles on Cu foils, which are responsible for incomplete graphene growth or nucleation of FLG islands. EP improves the reproducibility but FLG inclusions are still found in the monolayer graphene films, often associated with nanoparticles.

We demonstrate that reproducible growth of exclusively SLG is achieved by coating a thin W layer on the backside of Cu foils. The W layer completely blocks C diffusion from the backside towards the front surface of the Cu foil. These results substantiate that the morphology of the upper Cu surface is not the determining factor in the growth of FLG, even though FLG nucleate at particulate locations. Therefore, the tight control over the hydrocarbon flow usually needed during the CVD protocol to achieve purely SLG with full coverage can be relaxed. At the same time, the possibility of increasing the hydrocarbon flow enables the drastic diminution of the growth procedure duration. That leeway on the growth conditions (both in terms of hydrocarbon flow and growth duration) constitutes a major contribution in the perspective of industrializing the production of SLG.

## Methods

Details about the experimental techniques can be found in the SI.

**Pre-growth Cu foil treatment**

We use a single 30×30 cm$^2$ Alfa Aesar (AA) Cu foil (reference number #46365: 25 µm-thick, purity 99.8%, annealed, uncoated) for all the CVD growth experiments, from which we either cut 2×8 cm$^2$ pieces if subjected to EP or 1×1 cm$^2$ pieces otherwise. The Cu pieces are sonicated, maintained in a vertical position, in a mixture of 60 ml of deionized water (DIW) and 2 ml of glacial acetic acid (GAA) for 5 min, then rinsed in DIW for 2 min still in vertical position, rinsed in isopropanol for 2 min, and finally blow-dried with nitrogen. This cleaning method is referred to as the "standard cleaning".

Before EP, the Cu pieces are cleaned in the same way as described above. EP is performed by reproducing the experimental setup and conditions proposed in Ref. [12], with some adaptations. More specifically, we use a Coplin staining jar as a container. The Cu foil, fixed to a glass slide for easy handling (the grooves in the Coplin jar exactly match the size of the glass slide), is used as anode and a *circa* 1-mm-thick Cu plate (of the same size as the Cu foil) is used as cathode. Both electrodes are connected to the power supply using crocodile clips. It is important that the Cu foil is flat and parallel to the thicker Cu electrode to achieve reproducible, uniform EP. A constant voltage of 7 V is applied between the two electrodes for 60 s (inter-electrode distance: ~5 cm). The electrolyte solution is a mixture of 25 ml of DIW, 12.5 ml of phosphoric acid, 12.5 ml of ethanol, 2.5 ml of isopropanol, and 0.4 g of urea. After EP, the Cu foil on the glass slide is transferred for rinsing to a second Coplin jar containing DIW, and sonicated for 2 min. Finally, it is stored in ethanol.

**Graphene growth**

The samples are first laid on a flat fused silica boat and inserted into a tubular fused silica reactor at room temperature. An Ar flow of 2000 sccm is admitted for 15 min in the tube after sealing (purge step). Meanwhile, the hotwall furnace is pre-heated to 1050 °C. Next, the fused silica tube is introduced into the furnace and the Ar flow is reduced to 500 sccm. The sample is then exposed to Ar alone during 15 min and is oxidized at the surface due to residual oxidizing impurities. Thereafter, the Cu foil's surface is reduced for 45 min with the addition of 20 sccm of $H_2$. Afterwards, dilute methane ($CH_4$; 5% in 95% of Ar) is injected to grow graphene. For all the samples without W on the backside, the dilute $CH_4$ flow is set to 0.5 sccm, while it is variable otherwise (with a maximum of 3 sccm). The reactor is extracted 1 h later from the furnace after graphene growth and left to cool down naturally in the same gas mixture. The conditions corresponding to a dilute $CH_4$ flow of 0.5 sccm and a growth duration of 1 h are referred to as the "standard conditions" in the following. During the whole growth procedure, the reactor is maintained at atmospheric pressure (no pumping equipment connected to the system, which is called "vacuum-free"). For more details, we refer the reader to our previous publication.[28]

**Graphene transfer**

Graphene is transferred onto $SiO_2$/Si substrates by the widely used method based on poly(methyl methacrylate) (PMMA). After PMMA coating and baking at 110 °C, graphene grown on the backside of the Cu foil is removed by O plasma. Cu is etched in an ammonium persulfate solution. The PMMA/graphene stack is next rinsed thoroughly in DIW and fished on the SiO$_2$/Si support. The sample is left to dry overnight, baked at 120 °C for 1 h, and, finally, PMMA is stripped with acetone.

**W deposition**

W (50 nm; 99.95% purity) is coated on the backside of the Cu pieces by magnetron sputtering with a deposition pressure of $10^{-2}$ mbar (base pressure = $10^{-4}$ mbar) and Ar as sputtering gas. The deposited thickness is controlled by a quartz balance next to the sample.

**Acknowledgement**

The authors acknowledge C. Charlier for his help during the experiments. The research leading to this work received funding from the European Union H2020 program "Graphene Driven Revolutions in ICT and Beyond" (Graphene Flagship), project No 649953, and H2020 RISE project No 734164 "Graphene 3D". This work also made use of resources of the Electron Microscopy Service ('Plateforme Technologique Morphologie – Imagerie') and of the SIAM platform at the University of Namur. This work is also supported by the KU Leuven Internal Research Fund C14/17/080.

**Supporting Information Available:** 1) Reproducibility issues, 2) electropolishing, 3) Raman spectroscopy and optical contrast characterization of the samples, 4) additional experimental details. This material is available free of charge *via* the Internet at http://pubs.acs.org.

# Supporting information

**Restoring self-limited growth of single-layer graphene on copper foil via backside coating**

Nicolas Reckinger, Marcello Casa, Jeroen E. Scheerder, Wout Keijers, Matthieu Paillet, Jean-Roch Huntzinger, Emile Haye, Alexandre Felten, Joris Van de Vondel, Maria Sarno, Luc Henrard, and Jean-François Colomer

# Table of contents

# 1) Reproducibility issues

As-received copper foils must be processed before graphene growth. Notably, they are often covered with thin metallic oxide anticorrosion coatings.[1,2,3] Several groups also have reported the presence of particles of <u>endogenous</u> origin (<u>copper particles</u> formed during hydrogen annealing at elevated temperature[4,5] or resulting from the reduction of copper oxide particles formed during annealing without hydrogen,[6,7] <u>impurities</u> arising from segregation from the bulk of the copper foil,[8,9] or already present on the as-received surface[7,10,11,12] (see Table S1 for reported compositions)) or <u>exogenous</u> origin (contaminants from the quartz tube[13,14,15,16] or back streaming from downstream system components[17]). These particles promote few-layer graphene (FLG) nucleation,[4,10,12] prevent full-coverage graphene growth[12] or cause the etching of graphene.[16]

| Reference | Elements detected | Copper foil | Detection method |
|---|---|---|---|
| [2] | Ca, P, Cr | Alfa Aesar #13382 | XPS |
|  | Si | Alfa Aesar #10950 | XPS |
|  | Si | Alfa Aesar #42972 | XPS |
| [7] | Ca, P, Cr, N, Cl, O | Alfa Aesar #13382 | XPS |
| [10] | Si, Ca, Pt, Ru, Ce | Alfa Aesar (25 µm, 99.8% purity) | EDX |
| [12] | Si, Fe, Al, Ca, C, O | Not mentioned | EDX |

**Table S1:** Composition of impurities located on as-received copper foils reported in the literature.

We investigate here the cleaning of the copper foils to remove the impurity particles and/or the coating. Many different pretreatment techniques have been proposed such as: (1) chemical treatment in various liquids such as acetic acid,[2,18,19,20,21,22,23] solvents,[24] water,[18,25] inorganic acids (dilute $HNO_3$,[11,18] dilute HCl[11,26,27]), $FeCl_3$,[6,18] Cr or Ni etchants;[11] (2) electropolishing (EP)[3,18,19,20] and (3) chemical mechanical polishing.[4] More particularly, the popular removal of superficial copper oxide with acetic acid was shown to be very effective.[28] In the perspective of large-scale production, it also presents the distinctive asset not to involve complicated treatments or hazardous chemicals. Finally, some research groups even produced their own copper foils to have a tighter control on purity and roughness.[17,29]

We analyze the surface composition of as-received copper substrates by X-ray photoelectron spectroscopy (XPS) to assess the presence of contaminants. Two as-received samples (1×1 cm$^2$) from two different AA #46365 30×30 cm$^2$ batches are analyzed at three distinct randomly chosen spots. The concentration of the detected elements is determined from the survey scan (see Figure S1a). Besides copper, the main detected elements are oxygen (O 1$s$) and carbon (C 1$s$), corresponding to organic contamination. Chromium (Cr 2$p$; ~2%) is also unambiguously identified on both copper foils. Surprisingly, it indicates the presence of a layer of chromium oxide (the anticorrosion coating mentioned above), even though the AA #46365 copper foils are supposed to be coating-free (see Figure S1b displaying the high-resolution Cr core level spectra of three copper samples: as-received, reference (pure copper ingot), after growth). Other peaks are attributed to calcium (Ca 2*p*; ~6%) and phosphorus (P 2*p*; ~7%). The concentrations in the different elements can be found in Table S2. Besides it is worth noting that Murdock *et al.* have identified that the surface composition can vary from point to point on the same 30×30 cm$^2$ copper foil.[10] In our previous work,[7] we have relied on a cleaning mixture composed of glacial acetic acid (GAA) and deionized water (DIW) to remove the calcium- and phosphorus-bearing impurities. Figure S1c evidences the complete absence of graphene after growth in the standard conditions (see Methods in the main text) on an as-received sample inspected by scanning electron microscopy (SEM). The inset to Figure S1c shows the corresponding C 1*s* core level spectrum compared to the typical C 1*s* core level spectrum of single-layer graphene grown by chemical vapor deposition (CVD) (cleaning performed in GAA+DIW). Still, this GAA+DIW treatment does not strip the chromium oxide layer. Therefore, it is irrelevant to remove the superficial copper oxide layer, even though GAA is widely used for that purpose in the literature on coated copper foils (like the mainstream AA 13382 foils). However, even though the chromium oxide layer is not removed after the GAA+DIW treatment, it is evaporated during the pre-growth annealing.[7] In conclusion, the analysis of the composition of the copper foil after graphene growth indeed confirms that, within the detection limit of XPS, the contamination layers are removed (see Figure S1a).

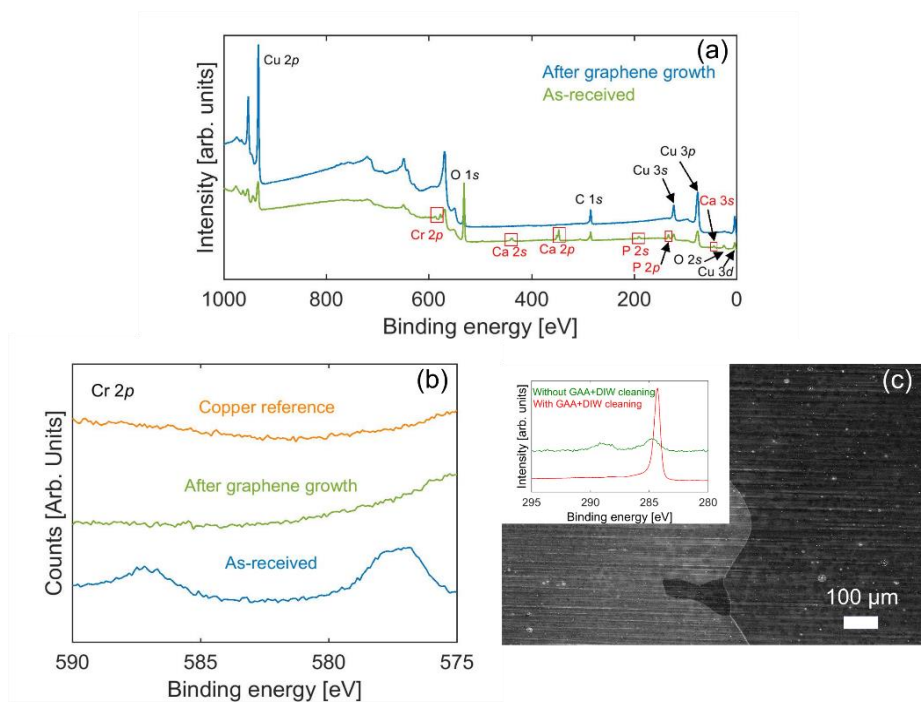

**Figure S1**: (a) X-ray photoelectron spectroscopy survey scan of the as-received and the graphene/copper samples. (b) Cr 2*p* core level spectra for the as-received, the graphene/copper, and the reference (pure copper ingot) copper pieces, respectively. (c) Scanning electron microscopy image of a post-growth copper piece without any pre-growth treatment displaying two large copper grains. Inset: C 1*s* core level spectra of two post-growth copper piece: one cleaned in the mixture of glacial acetic acid and deionized water and the other one without any pre-growth cleaning.

Despite this efficient treatment, the large-scale inspection of more than 25 1×1 cm$^2$ graphene/copper pieces coming from all over the surface of the 30×30 cm$^2$ batch by low-magnification SEM of graphene grown under the standard conditions manifests poor reproducibility both in terms of coverage and FLG formation (see Figure S2a,b and Figure S3a-c). Only 50% of the samples are fully covered, with a highly variable amount of FLG. One could naively argue that higher coverage can be reached by increasing the dilute methane flow, but this is done at the expense of the FLG concentration. In the following text, we only provide data for a few representative samples.

| Sample type | Point | Atomic composition [%] | | | | | |
|---|---|---|---|---|---|---|---|
| | | O 1$s$ | Cu 2$p$ | C 1$s$ | Ca 2$p$ | P 2$p$ | Cr 2$p$ |
| As- received batch #1 | 1 | 51.34 | 11.79 | 23.82 | 5.28 | 6.02 | 1.75 |
| | 2 | 50.36 | 11.33 | 25.02 | 5.27 | 6.46 | 1.56 |
| | 3 | 50.95 | 11.54 | 24.04 | 5.36 | 6.33 | 1.78 |
| As- received batch #2 | 1 | 54.3 | 12.44 | 17.73 | 6.21 | 7.18 | 2.14 |
| | 2 | 53.26 | 12.22 | 19.8 | 6.03 | 6.71 | 1.98 |
| | 3 | 54.1 | 10.57 | 19.22 | 6.5 | 7.24 | 2.36 |
| After graphene growth | 1 | 23.04 | 43.42 | 34.54 | X | X | X |
| | 2 | 24.62 | 41.75 | 33.63 | X | X | X |
| | 3 | 23.09 | 47.38 | 29.53 | X | X | X |

**Table S2:** Surface composition, obtained from X-ray photoelectron spectroscopy analyses, of two as-received Alfa Aesar #46365 copper samples from two different batches, and one sample after graphene growth under the standard conditions.

A first cause for this lack of reproducibility and homogeneity is the incomplete surface contamination removal. Upon closer inspection by SEM (see Figure S2d), we can indeed observe that micro- and nanoparticles still reside on the copper foil after graphene growth. Note that XPS is not sensitive enough to detect them. All these particles (present on the copper foil at the same location before the CVD growth) cause discontinuities in the graphene film because they hinder graphene growth at their location (see Figure S2c and Figure S3a), and in addition some of them are the source of uncontrolled FLG island nucleation (see Figure S2d). For instance, as observed in Figure S2c, the graphene coverage on this copper piece is almost complete except for a small uncovered area. By zooming on this area (see Figure S2d), we can observe that it is covered by a line of particles of various sizes. Energy-dispersive X-ray spectroscopy can be used for a targeted analysis of these particles but it must be conducted on large enough (micron-sized) particles since prohibitively high acquisitions times are necessary to resolve nanometer-sized particles. The composition of two randomly selected areas in the graphene-free zone reveals the presence of aluminum oxide, silicon oxide, and carbon, as already reported in Ref. [10] (see Figure S4 and Table S3 for the precise composition). Further examples of the heterogeneity can be found in Figure S3d-j. It reveals the surface of several copper specimens after annealing in argon, and after annealing in argon followed by reduction in hydrogen (i.e. the morphology of the copper foil just before graphene growth).

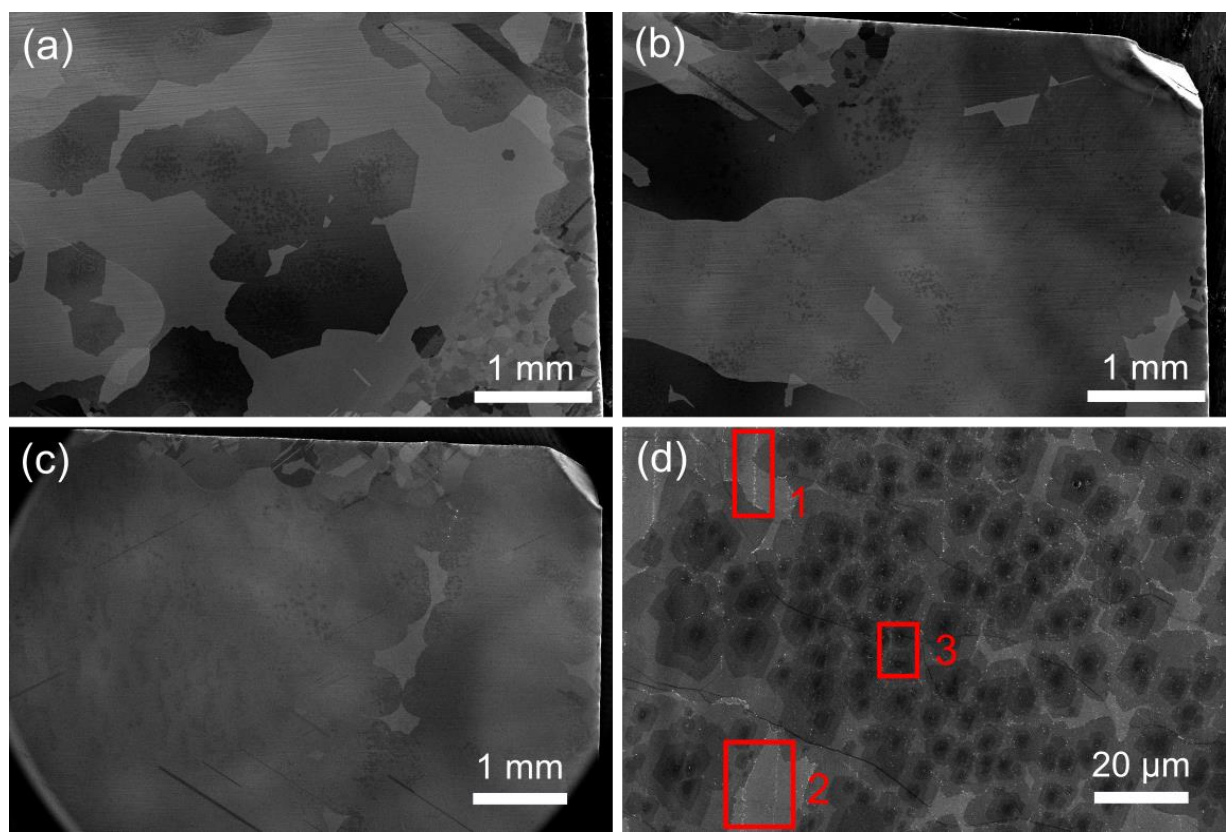

**Figure S2**: (a-c) Low-magnification scanning electron microscopy picture of two samples grown sequentially in the standard conditions. (d) Zoom-in on an area with a high-density of few-layer graphene islands. The red rectangles 1 and 2 illustrate the presence of white particles along the rolling striations and the rectangle 3 the nucleation of few-layer graphene on white particles.

(a)

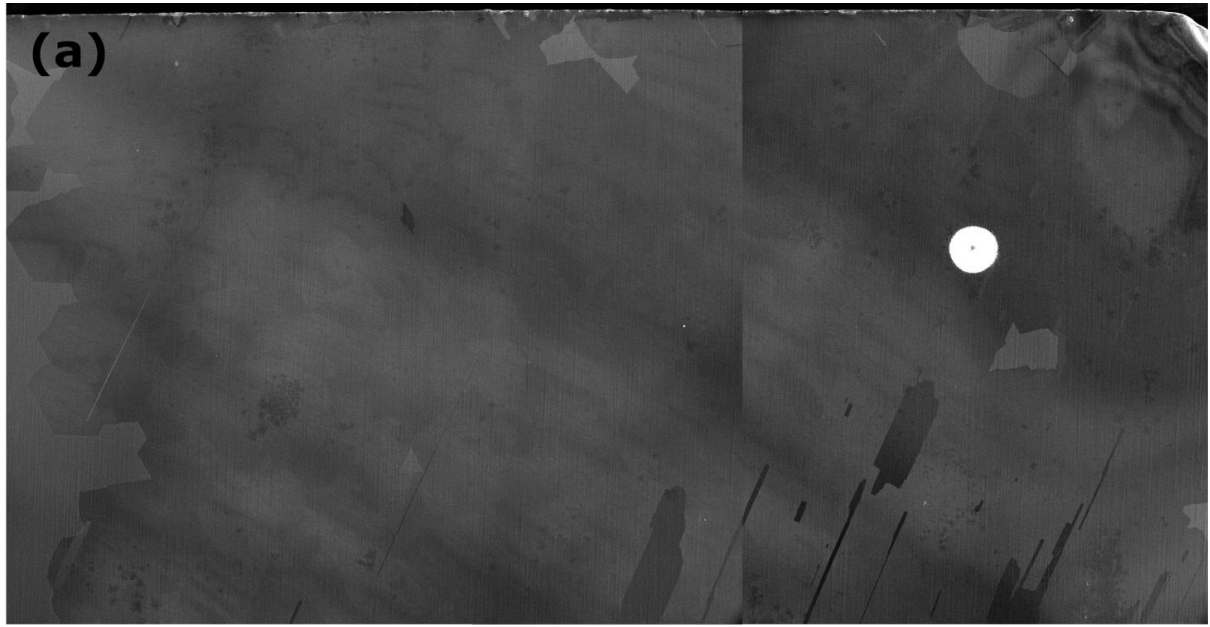

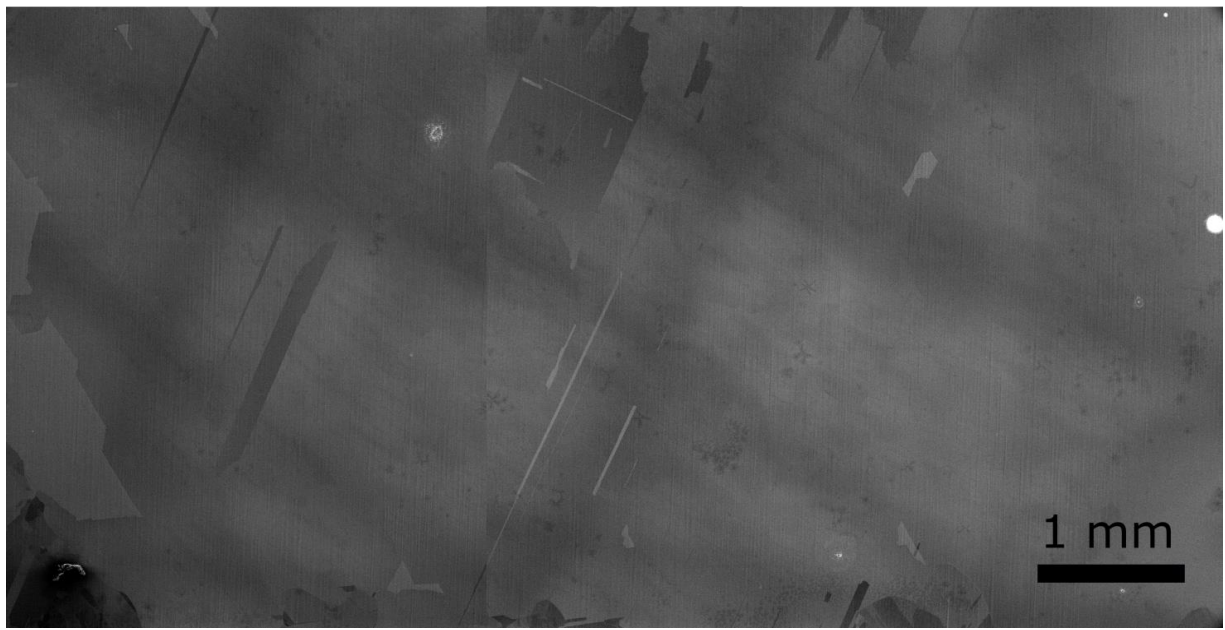

1 mm

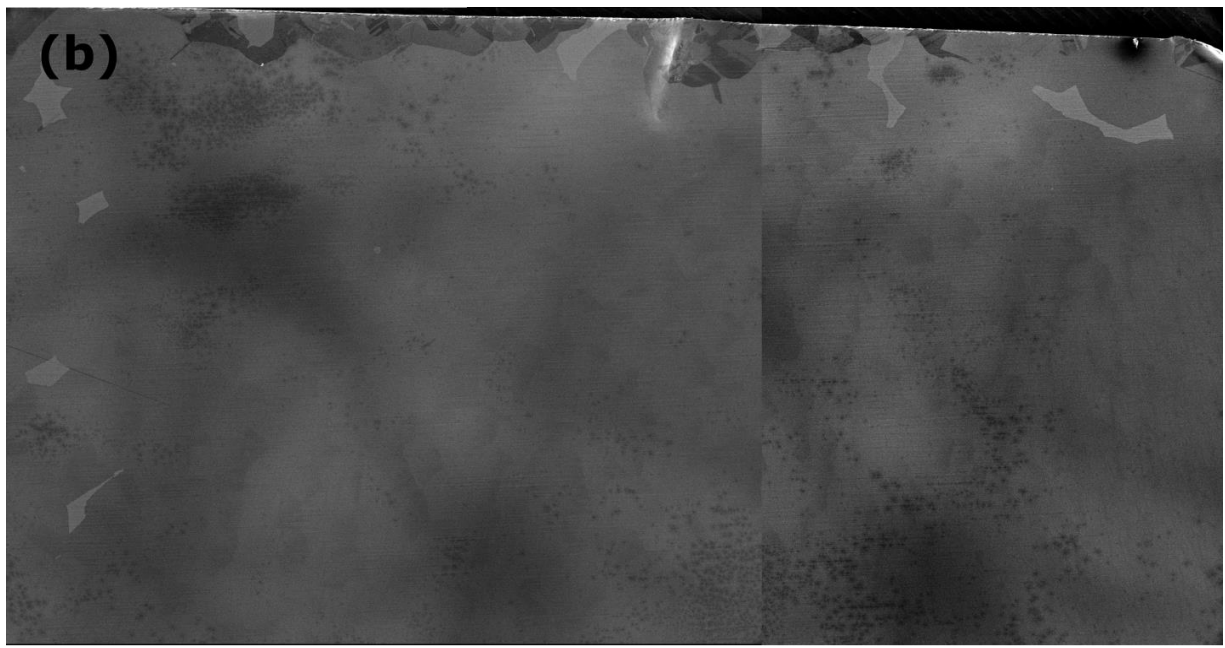
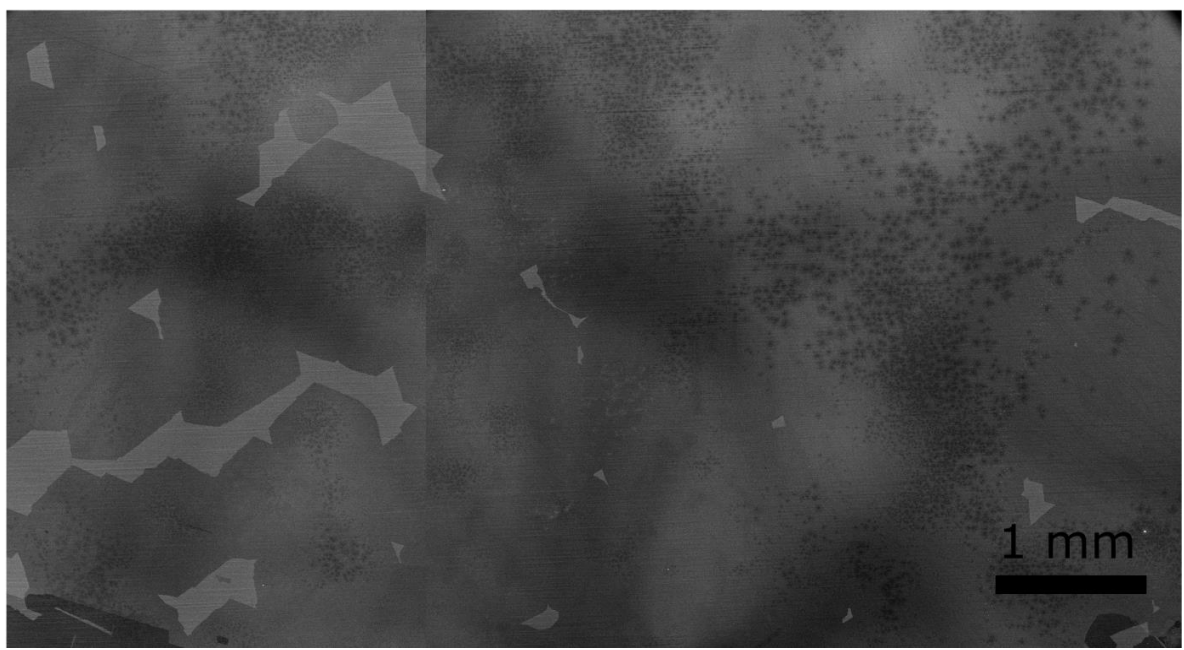

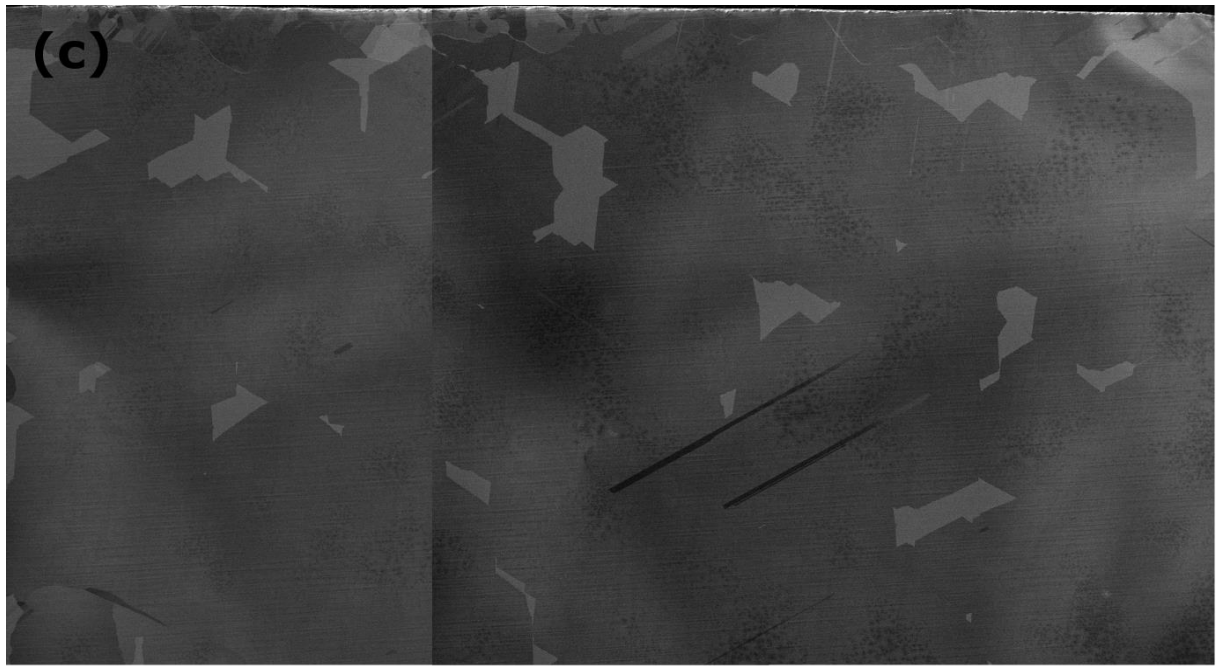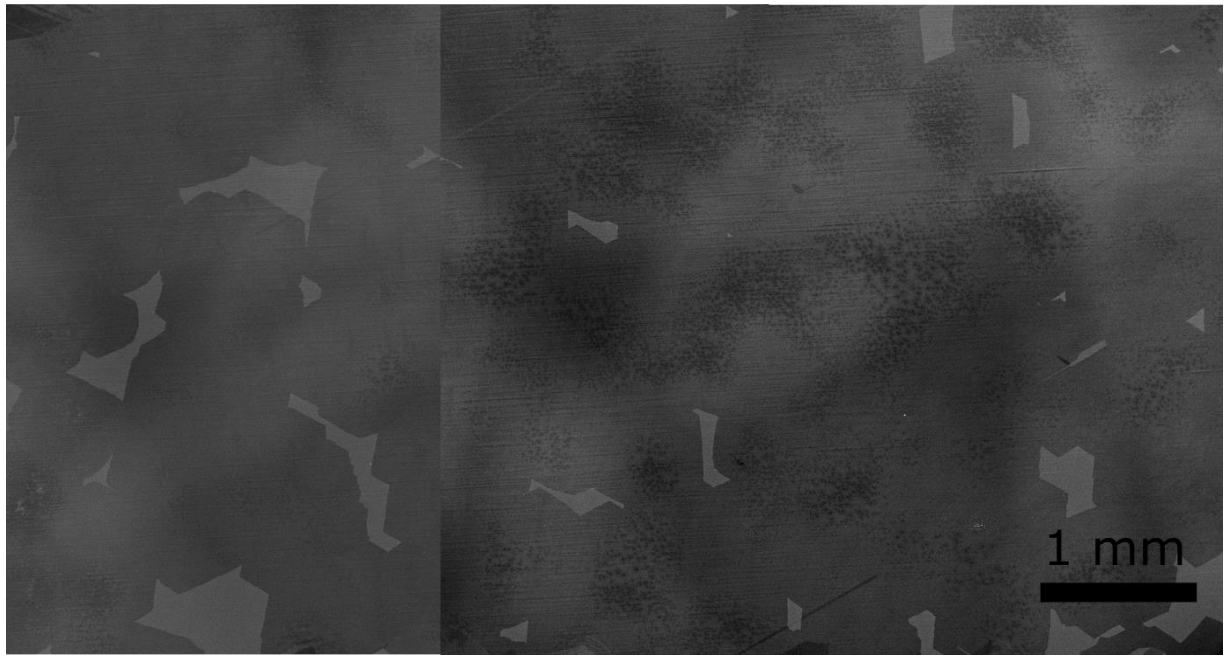

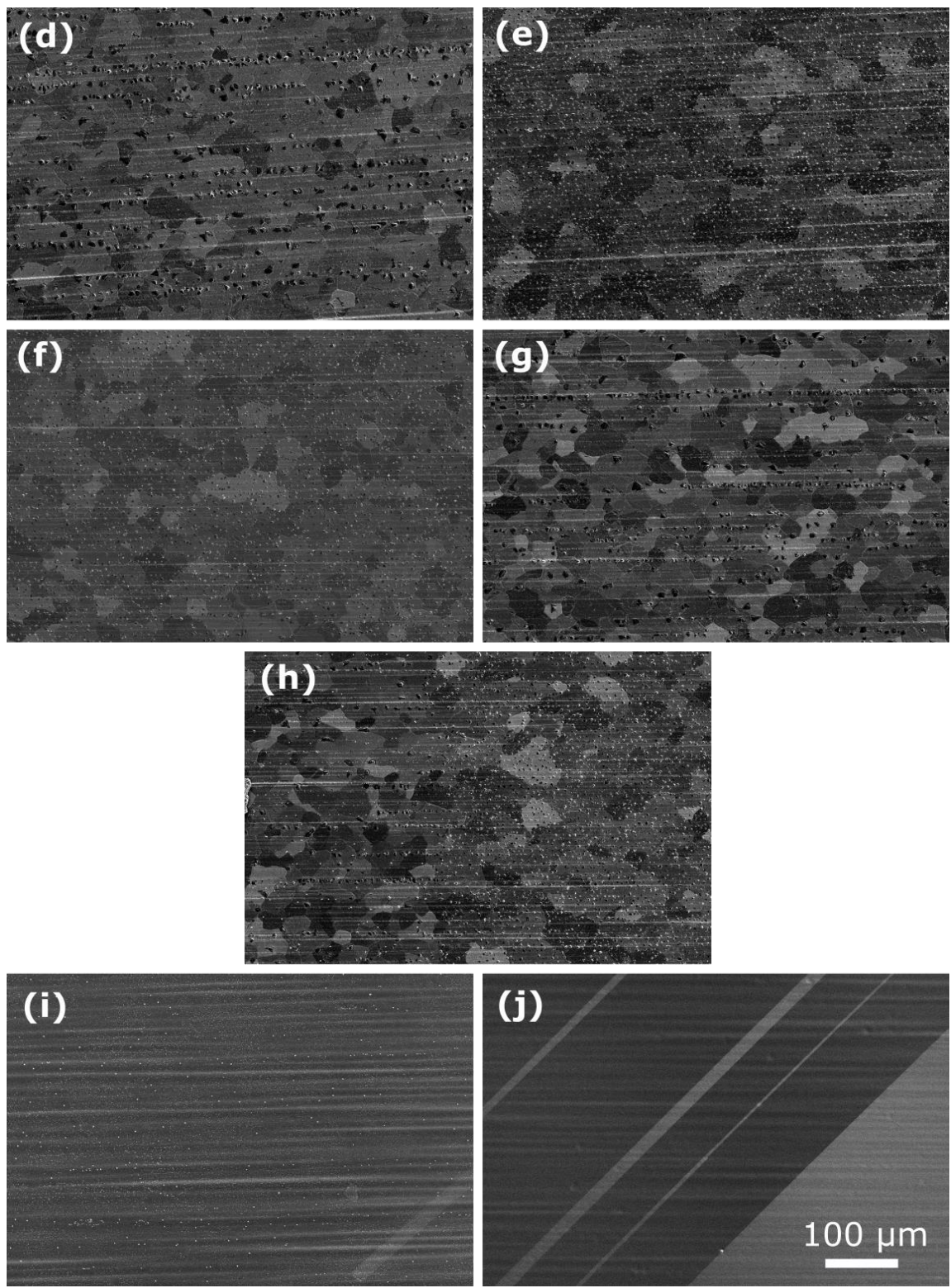

**Figure S3**: (a)-(c) Low-magnification scanning electron microscopy pictures of the whole surface of three samples grown altogether in the standard conditions. The bottom and top halves in each figure are formed by stitching two pictures together. Higher magnification scanning electron microscopy pictures of five samples just after the 15 min oxidation under argon alone (d)-(h) and two additional samples just after oxidation (15 min) and reduction (45 min) (i-j). The scale bar is identical for images from (d) to (j).

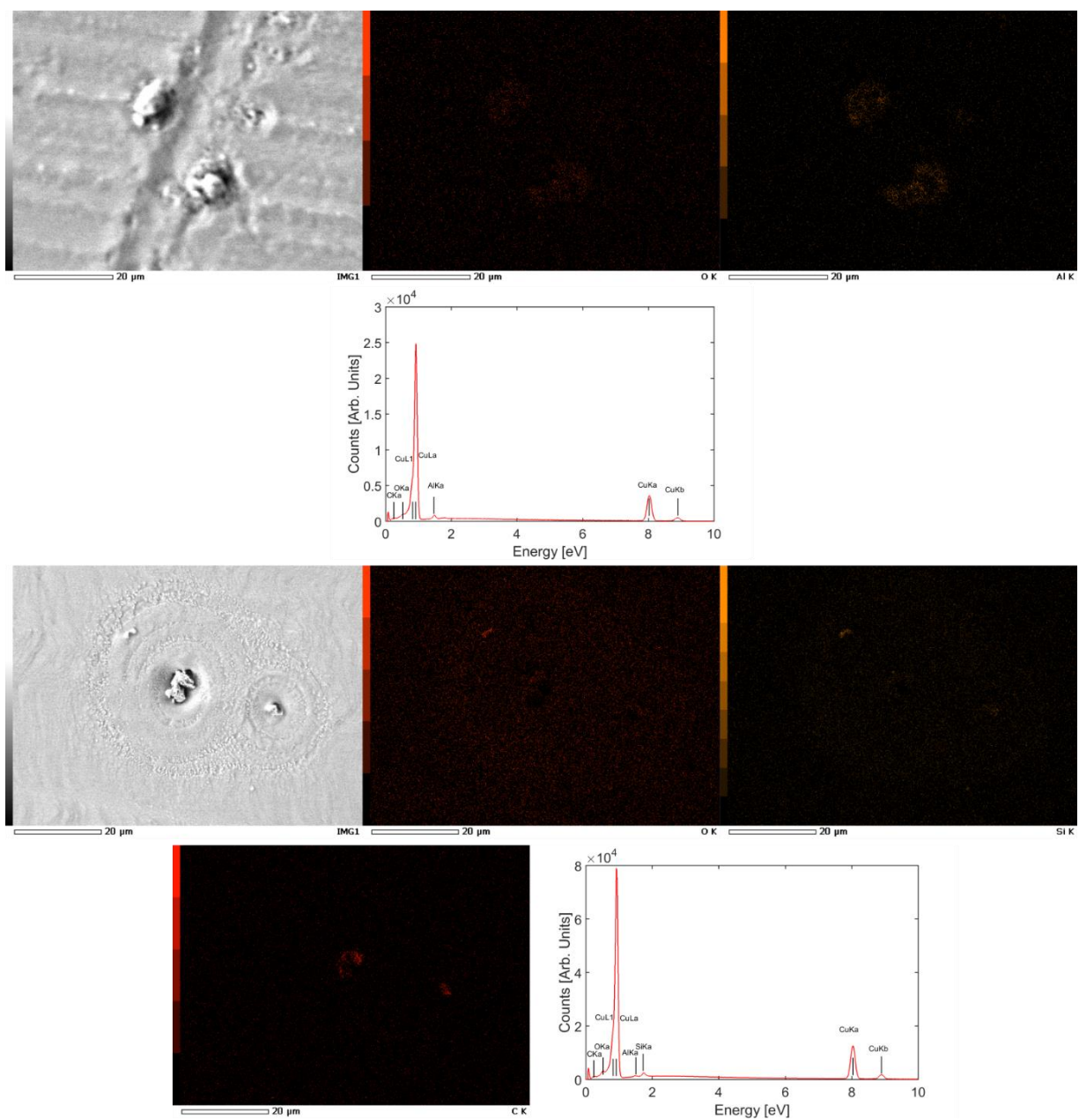

**Figure S4:** Energy-dispersive X-ray spectroscopy elemental mappings and spectra of two graphene-free areas for a graphene/copper sample grown under the standard conditions, exhibiting micrometer-sized particles.

| Atomic composition [%] | | | | | |
|---|---|---|---|---|---|
| Area | C K | O K | Al K | Si K | Cu K |
| 1 | 10.34 | 2.49 | 2.93 | X | 84.24 |
| 2 | 9.64 | 4.29 | 0.6 | 1.46 | 84.02 |

**Table S3:** Elemental composition of the two areas investigated in Figure S4.

In order to reduce the presence of these particles, we test various other techniques found in the literature: (1) chemical cleanings/etchings with various chemicals such as dilute nitric acid (5% in water), hydrochloric acid dip followed by potassium hydroxide dip, with an intermediate rinsing in DIW (same conditions as in Ref. [10]), ammonium persulfate (0.3 M), GAA+DIW followed by immersion in Transene chromium etchant and (2) growth with a copper piece stacked over a nickel foil after GAA+DIW treatment. Unfortunately, in our conditions, none of these methods give satisfactory results to jointly realize uniformity and repeatable coverage of graphene (see Figure S5). For surface copper etching, the treatments increase the roughness,[11] providing more defective regions prone to nanoparticle formation during the oxidative pre-growth annealing (see just below).

Alternatively, in addition to impurities, copper nanoparticles can appear during the CVD process (before the graphene synthesis) owing to the pre-growth annealing under argon (and residual oxidizing gas impurities), after reduction in hydrogen.[6,7] Defected zones such as rolling striations or scratches are more specifically susceptible to oxidation, as we can see in Figure S2d, where white particles are seen all along the striations. Figure S2d clearly displays that these particles are the source of unwanted FLG nucleation. The lack of control over the amount of residual gaseous oxidizing impurities in the reactor of our vacuum-free setup between each CVD procedure is an additional source of variability (as testified by the pictures in Figure S3d-j). To investigate the impact of the quantity of residual oxidizing impurities on the final result, we reduce the amount of argon supply during the synthesis (150 or 300 sccm during all the process, instead of 2000 sccm for 15 min then 500 sccm for the remainder of the process, all other conditions being standard). In this way, the atmospheric air contained in the reactor (left open between two CVD syntheses) is less diluted when mixed with argon during the purge step, augmenting the partial pressures in unwanted oxidizing species such as oxygen and water. As seen in Figure S6, large patches of FLG spotted with thicker, much smaller FLG flakes appear as the argon flow drops.

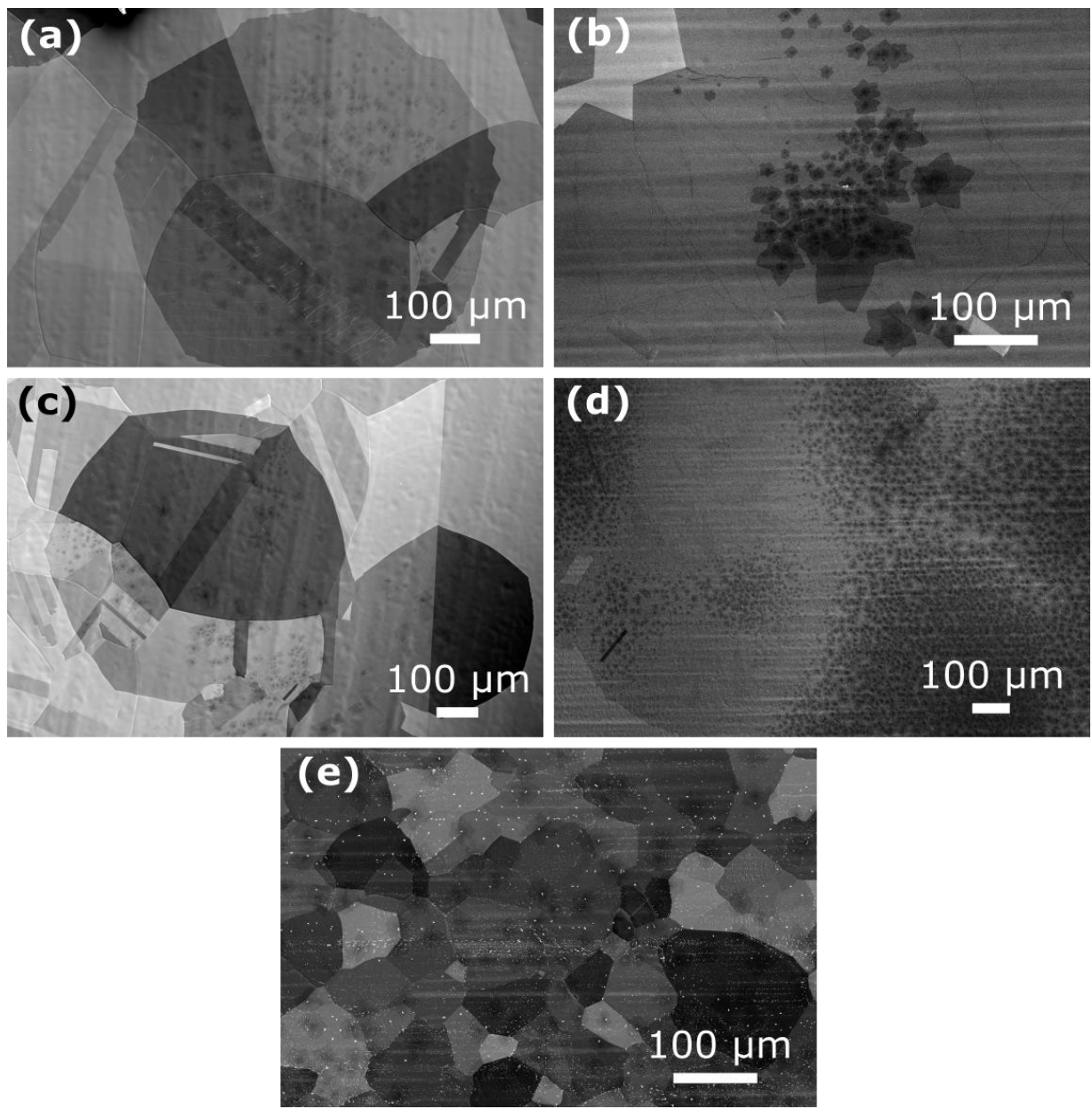

**Figure S5:** Scanning electron microscopy images of several samples grown under the standard conditions, subjected to diverse pre-growth treatments: (a) dilute nitric acid, (b) hydrochloric acid dip followed by potassium hydroxide dip, with an intermediate rinsing in deionized water (c) ammonium persulfate, (d) mixture of glacial acetic acid and deionized water followed by immersion in Transene chromium etchant. (e) Graphene growth (under the standard conditions) with the copper sample lying over a nickel foil, after glacial acetic acid and deionized water pre-growth treatment. None of these techniques lead to reproducibility in terms of coverage and homogeneity simultaneously. It is however worth mentioning that treatment (b) results in full graphene coverage, but with areas comprising few-layer graphene islands.

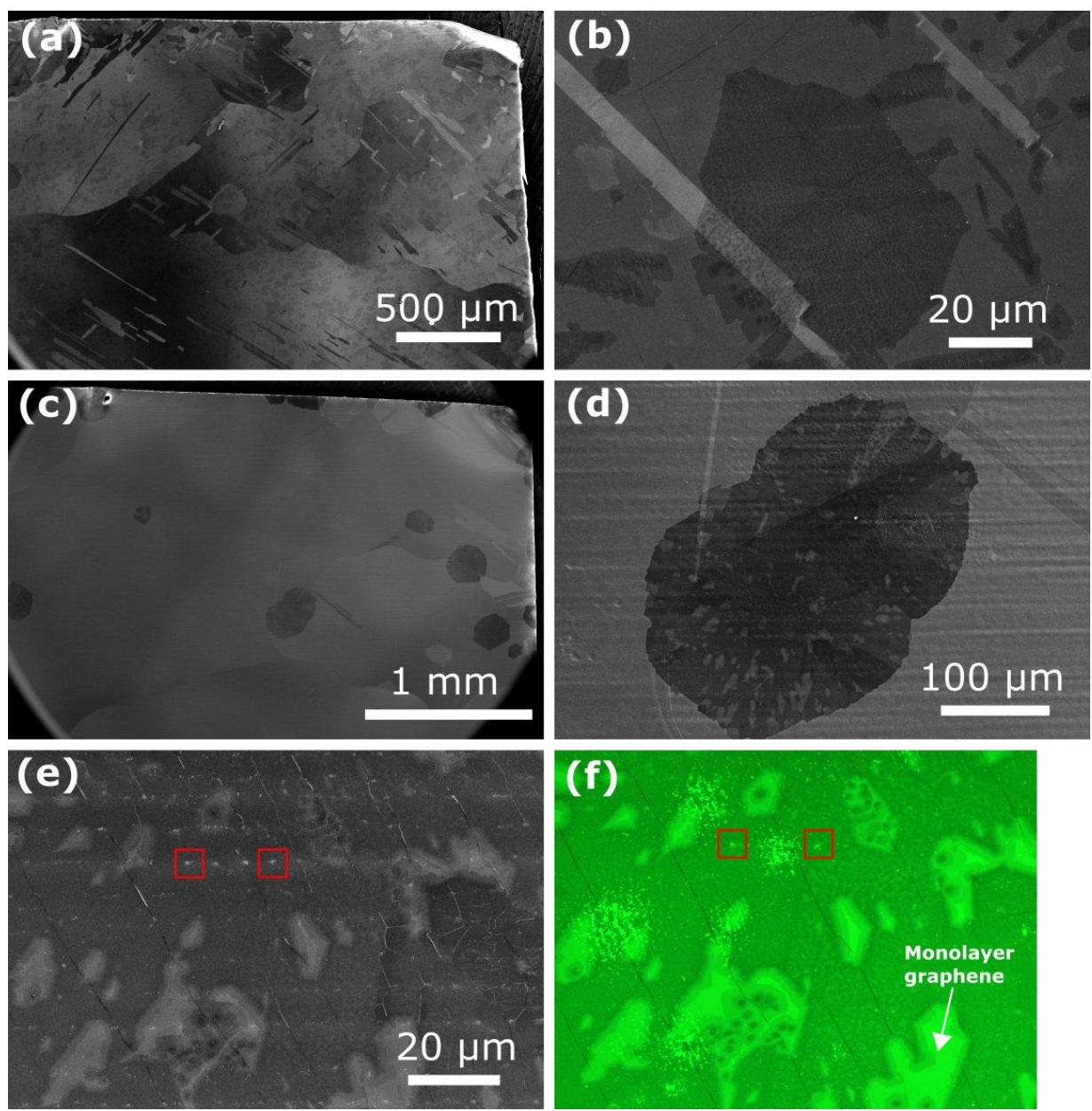

**Figure S6:** Scanning electron microscopy images of two graphene/copper samples grown under the standard conditions, except for the argon flow, which is decreased to 300 (a),(b) or 150 sccm (c)-(e), respectively. This argon flow drop has the effect of increasing the partial pressure in oxidizing impurities in the reactor, with the consequence of a strong increase in few-layer graphene areas. (f) Optical microscopy image (with a green filter to increase the graphene layer number contrast) of the same area as in panel (e) after transfer onto a 300-nm-thick silicon dioxide piece. Incidentally, another adverse side effect of the presence of contamination on the copper foil before graphene growth is that they are transferred with graphene on the target substrate (not to mention the impurities introduced by the transfer procedure itself).

## 2) Electropolishing

We turn our attention to a more radical treatment: EP (see the experimental section for more details). In Figure S7a-b, we show the same copper stripe (2×8 cm$^2$) before and after EP. The reproducibility of the technique in terms of graphene coverage is assessed by growing graphene (still in the standard conditions) on nine ~1×1 cm$^2$ copper pieces cut from three different EP batches. Figure S7c displays a photograph of the nine copper pieces after growth and oxidation at 150 °C for 5 min on a heating plate to reveal bare copper. It appears clearly that the EP process leads to very reproducible graphene coverage. The improvement in coverage control obtained for 9 samples synthesized in a row from different EP batches is an indication that EP, as expected, removes all the surface contamination on the copper specimens. However, it is not yet completely satisfying as far as uniformity is concerned, as exemplified by the picture shown in Figure S7d. We believe that this may be due to the poor control over the amount of residual gaseous oxidizing impurities during the hydrogen-free pre-growth annealing (see Figure S3d-j).

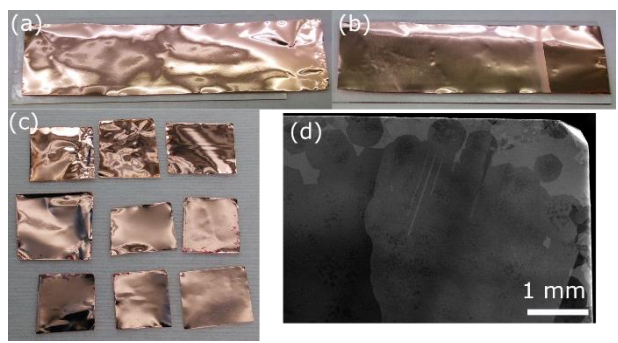

**Figure S7**: Photographs of the same copper stripe (a) before and (b) after electropolishing. (c) Photograph of nine electropolished copper specimens after graphene growth in the standard conditions. (d) Low-magnification scanning electron microscopy picture of one of the nine electropolished copper samples.

## 3) Raman spectroscopy and optical contrast characterization of the samples

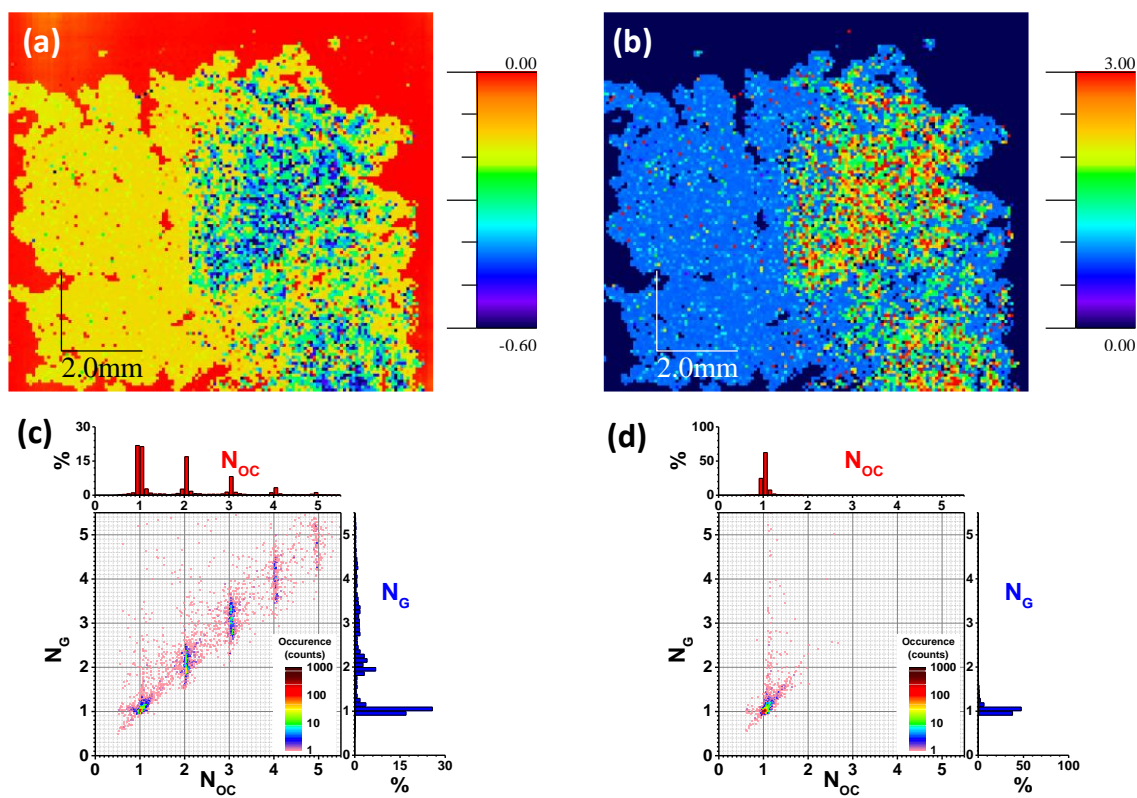

**Figure S8:** (a) Optical contrast and (b) $A_G^{norm}$ maps (≈1×1 cm² with a 70 µm *xy*-step) of the sample shown in Figure 1c (dilute methane flow of 0.6 sccm during 1 h). (c) and (d) 3D bivariate histogram (0.025 bin size) of $N_{OC}$ and $N_G$ derived from maps (a) and (b) using the expressions 3 and 4 from Ref. [30] (displayed at the end of this section). (c) corresponds to the full map while, for (d), only the left half of the sample is considered. The number of occurrences (frequency counts) is color-coded as shown on the plots. On top (resp. right hand side) are displayed the corresponding histograms of $N_{OC}$ (resp. $N_G$).

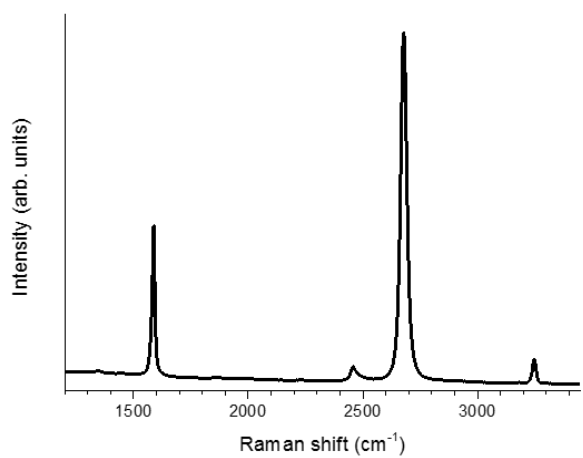

**Figure S9:** Average Raman spectrum of the left part of the sample shown in Figure 1c (dilute methane flow of 0.6 sccm during 1 h) corresponding to the points presented in Figure S8d.

S16

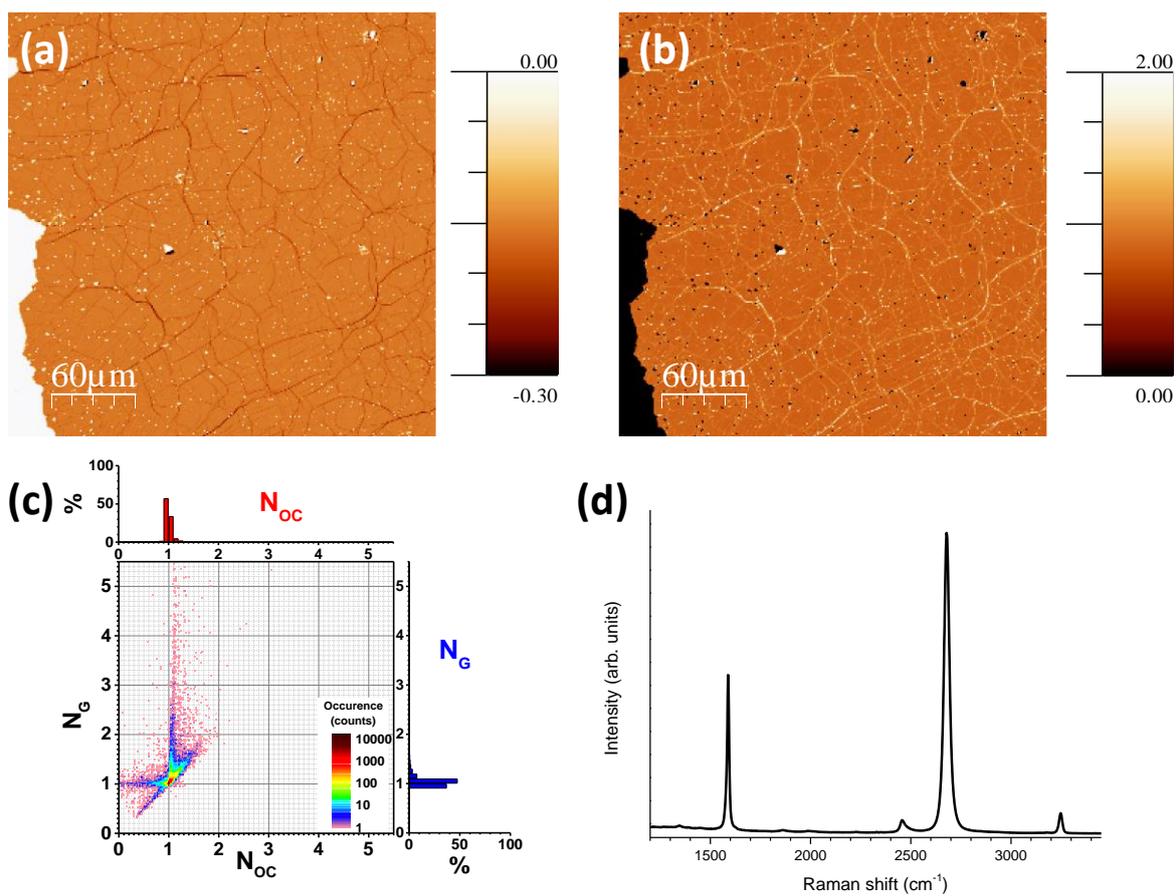

**Figure S10:** (a) Optical contrast and (b) $A_G^{norm}$ maps (300×300 µm² with a 1 µm *xy*-step) of the sample shown in Figure 1c (dilute methane flow of 0.6 sccm during 1 h). (c) 3D bivariate histogram (0.025 bin size) of $N_{OC}$ and $N_G$ derived from maps (a) and (b) using the expressions 3 and 4 from Ref. [30] (displayed at the end of this section). (d) Average Raman spectrum corresponding to the map displayed in (b).

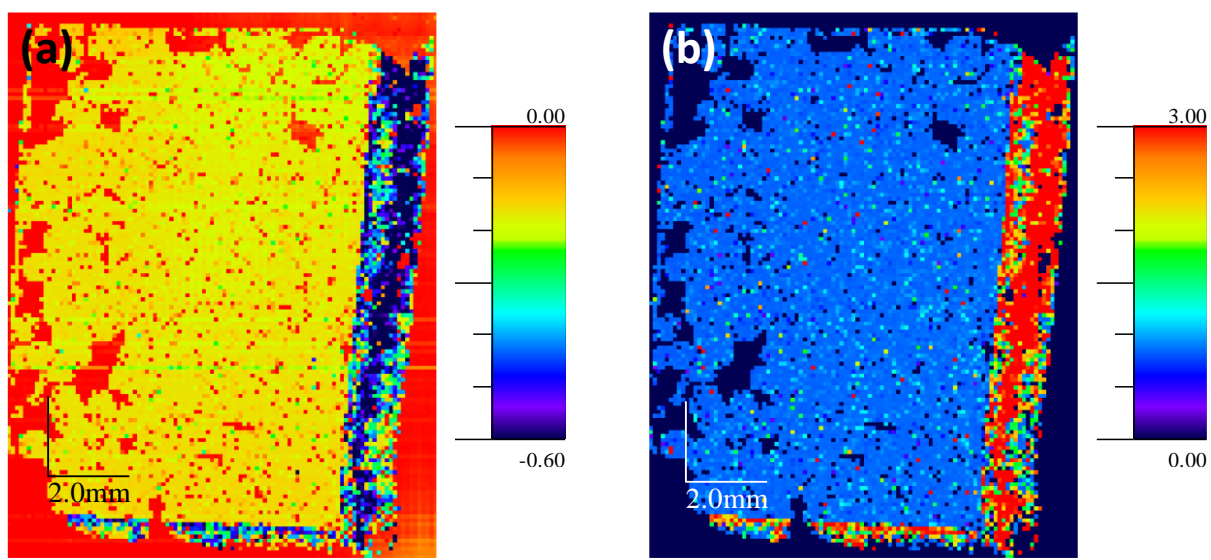

**Figure S11:** (a) Optical contrast and (b) $A_G^{norm}$ maps ($\approx$1.05×1.35 cm² with a 100 µm $xy$-step) of the sample shown in Figure 2 (dilute methane flow of 1.2 sccm during 1 h).

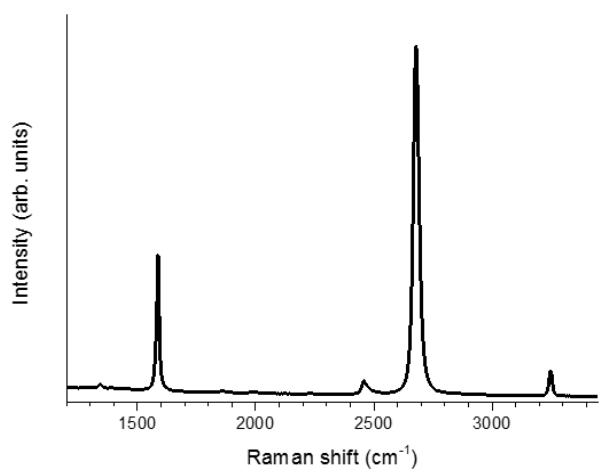

**Figure S12:** Average Raman spectrum of the sample shown in Figure 2 (dilute methane flow of 1.2 sccm during 1 h), corresponding to the region delimited by the dashed rectangle.

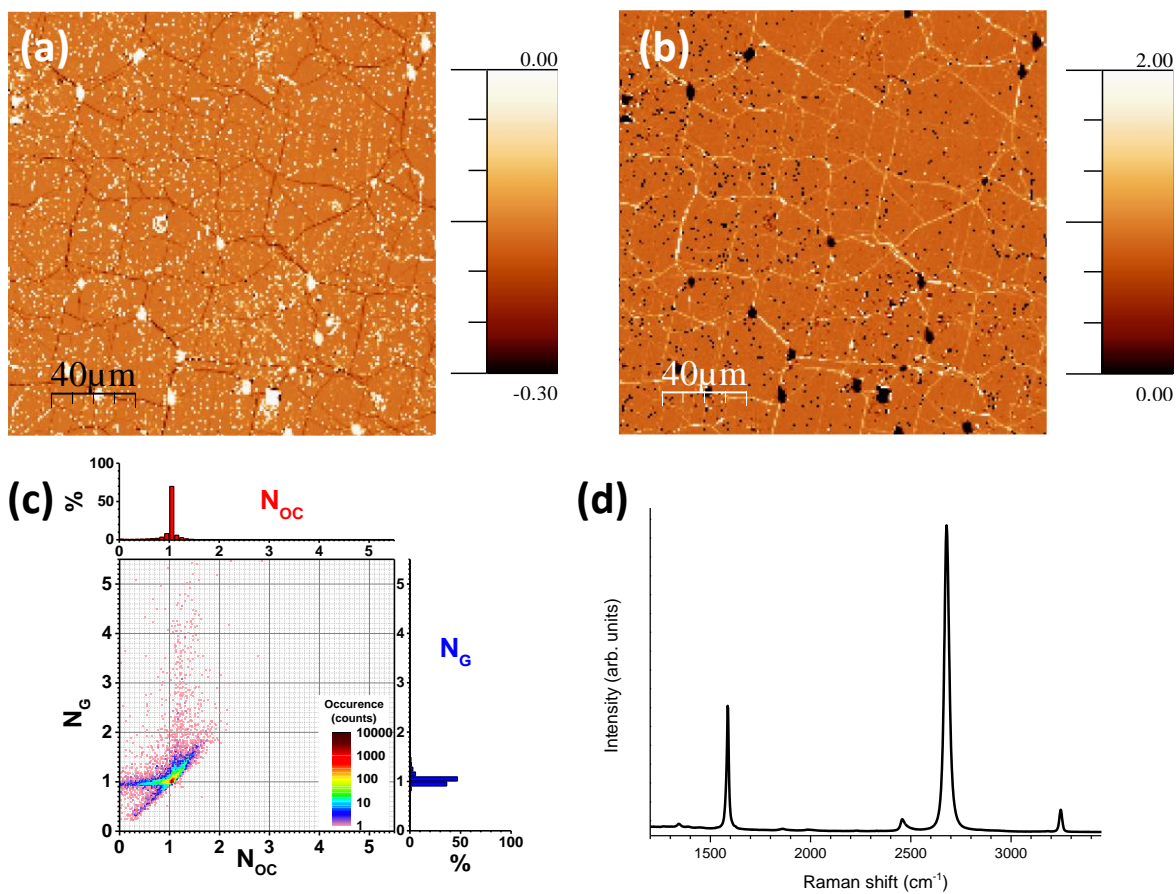

**Figure S13:** (a) Optical contrast and (b) $A_G^{norm}$ maps (200×200 µm² with a 1 µm *xy*-step) of the sample shown in Figure 2 (dilute methane flow of 1.2 sccm during 1 h). (c) 3D bivariate histogram (0.025 bin size) of $N_{OC}$ and $N_G$ derived from maps (a) and (b) using the expressions 3 and 4 from Ref. [30] (displayed at the end of this section). (d) Average Raman spectrum corresponding to the map displayed in (b).

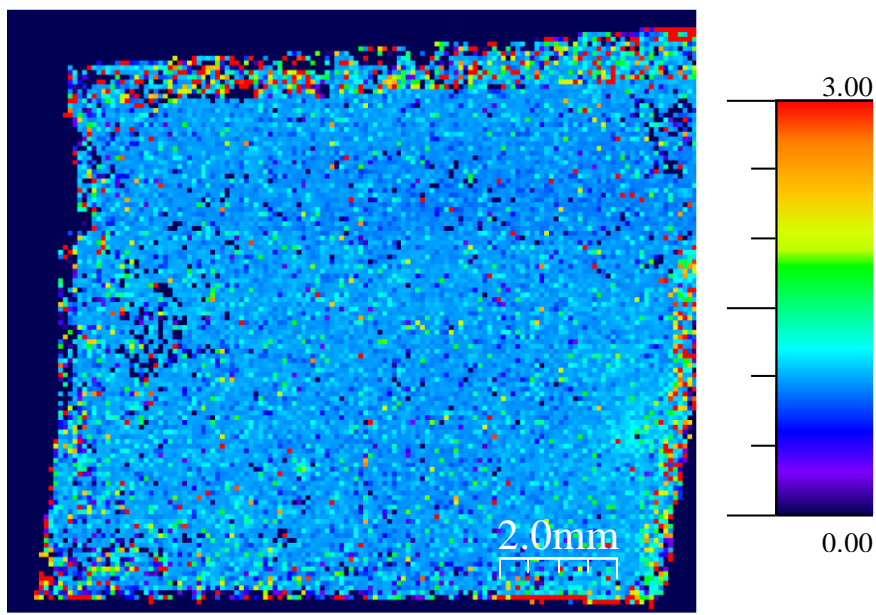

**Figure S14:** $N_G$ maps ($\approx$1×1.15 cm² with a 80 µm *xy*-step) of the sample grown with a dilute methane flow of 3 sccm during 5 min, derived from $A_G^{norm}$ (not shown) using the expression 3 from Ref. [30] (expression (1) given at the end of this section).

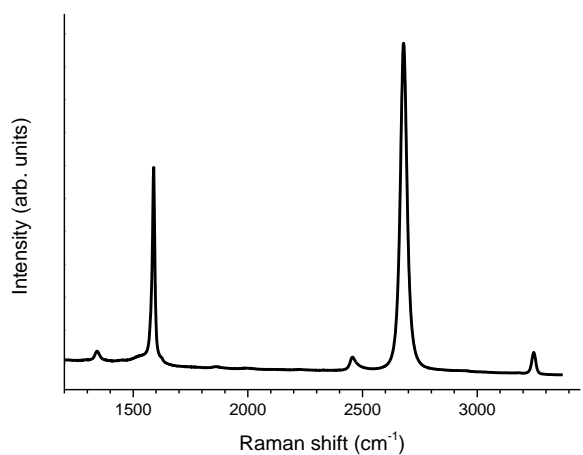

**Figure S15:** Average Raman spectrum of the sample grown with a dilute methane flow of 3 sccm during 5 min corresponding to the map of Figure S14.

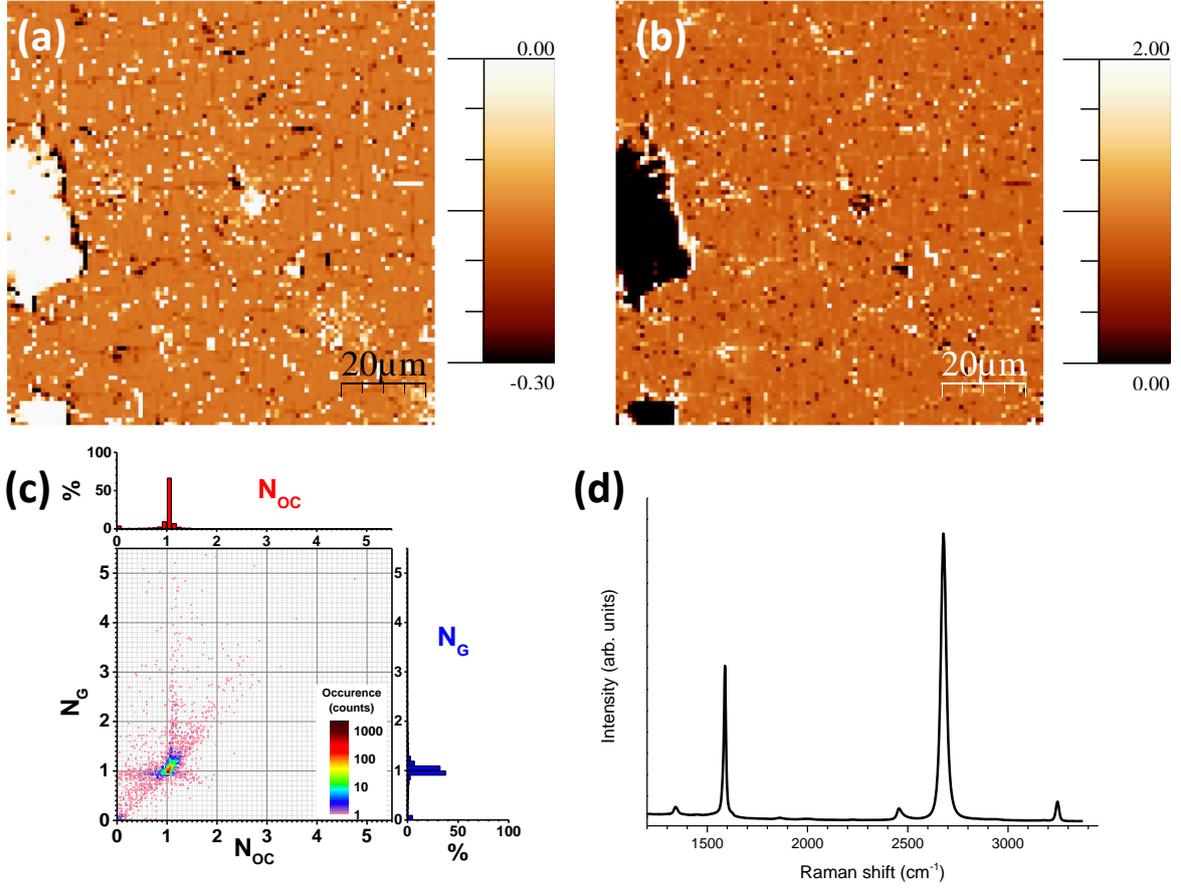

**Figure S16:** (a) Optical contrast and (b) $A_G^{norm}$ maps (100×100 µm² with a 1 µm *xy*-step) of the sample grown with a dilute methane flow of 3 sccm during 5 min. (c) 3D bivariate histogram (0.025 bin size) of $N_{OC}$ and $N_G$ derived from maps (a) and (b) using the expressions 3 and 4 from Ref. [30] (displayed at the end of this section). (d) Average Raman spectrum corresponding to the map displayed in (b).

## Expressions used to calculate the estimated numbers of layers $N_G$ and $N_{OC}$

$N_G$ and $N_{OC}$ (the number of layers estimated from $A_G^{Norm}$ and from the optical contrast (OC), respectively) are obtained using the relations found in Ref. [30]:

$$N_G = 1.05 \times A_G^{Norm} + 0.16 \times \left(A_G^{Norm}\right)^2 \qquad (1)$$

and

$$N_{OC} = -5.74 \times OC + 4.61 \times OC^2. \qquad (2)$$

## 4) Electrical characterization

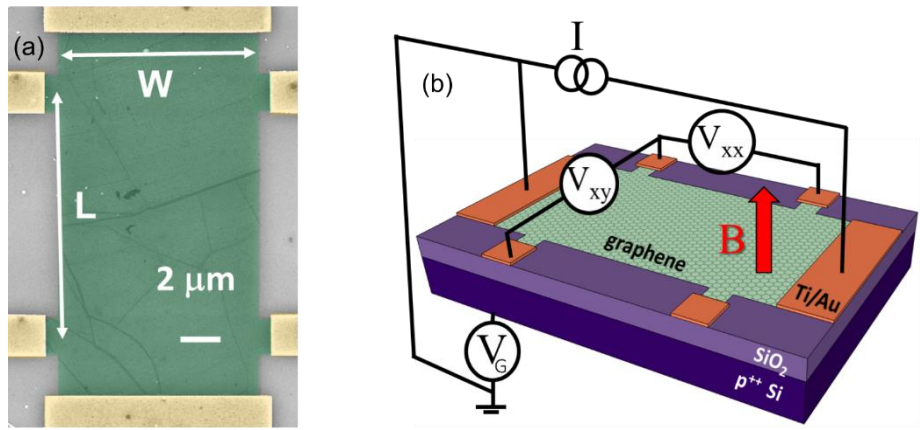

**Figure S17:** (a) False-colored scanning electron micrograph of a graphene field-effect transistor, with single-layer graphene (green) contacted by Ti/Au electrodes (yellow). (b) Schematic overview of the circuit for field-effect measurements. A current bias of 10 µA is applied to the sample, while the longitudinal voltage $V_{xx}$ and transversal (Hall) voltage $V_{xy}$ are measured as a function of the gate voltage $V_G$ and of the magnetic field $B$, applied perpendicular to the graphene plane.

# 5) Additional experimental details

**Scanning electron microscopy/Energy-dispersive X-ray spectroscopy**

The morphology and the size of the Cu grains and graphene domains are monitored with two different microscopes: a Jeol JSM-6010LV InTouchScope at low magnification (operated at an accelerating voltage of 5 kV and a spot size between 30 and 50, with a working distance of 25 mm to increase the field of view, in secondary electron mode) and a Jeol JSM-7500F at high resolution (operated at an accelerating voltage of 1 kV and an emission current of 5 µA, with a working distance of 3 mm, in secondary electron mode with a low gentle beam of 0.2 kV applied to the specimen). Energy-dispersive X-ray spectroscopy mapping is performed with the Jeol JSM-7500F at 15 kV with a probe current of 1 nA and a resolution of 512×384 px.

**X-ray photoelectron spectroscopy**

Two different spectrometers are used: a K-alpha and an Escalab250Xi from ThermoFischer Scientific. The spectra are recorded at constant pass energy (150 eV for survey; 30 eV for high-resolution spectra; 50 eV for depth profiling). A flood gun (low energy electrons and $Ar^+$ ions) is used during all the measurements. During the sputtering, the $Ar^+$ ion gun is operated at an accelerating voltage of 2 kV, with an erosion step of 100 s per cycle. The X-ray photoelectron spectroscopy data are treated with the Avantage software. High-resolution spectra are fitted by Gaussian-Lorentzian lineshapes with an Avantage "smart" background (i.e. a Shirley background in most cases, or a linear background in case the lineshape decreases with increasing binding energy). The diameter of the analyzed surface is 250 µm for the surface analysis and 400 µm for the depth profile.

**Raman spectroscopy setup**

Raman spectra are recorded using an Acton SP2500 spectrometer fitted with a Pylon CCD detector and a grating that enables the measurement of the full spectrum in the 1000−3000 $cm^{-1}$ range within a single acquisition (*i.e.* for a 532 nm laser, 600 grooves/mm grating corresponding to ~2 $cm^{-1}$ between each CCD pixel). The samples are excited with a 532 nm (2.33 eV) laser (Millennia Prime, Newport) through a 100× objective (Numerical Aperture 0.9) and 1 mW impinging on the sample. Optimized focus conditions are checked for each measurement. The samples are mounted on a three-axis piezoelectric stage (Physik Instrumente, 300 µm ranges in xyz) to ensure the precise positioning and focusing of the laser spot and a two-axis piezoelectric stage (Physik Instrumente, 25 mm ranges in xy) for large maps (*i.e.* larger than 300 µm). The laser power is continuously measured by a calibrated photodiode put behind the beam splitter which enables to correct the laser power fluctuations during the sample mapping. To perform simultaneously microreflection (OC) measurements and Raman spectroscopy, a low noise photodiode is placed on the path of the laser beam reflected by the edge filter located in front of the spectrometer's entrance slit. The laser OC, defined by OC = (R − Rs)/Rs, where R (resp.

Rs) is the reflected intensity of the 532 nm laser light measured at each point of the sample (resp. on the bare substrate). The whole experimental setup (spectrometer, piezoelectric stage, photodiodes…) is controlled by a dedicated, home-made Labview application. Graphite is used as the Raman intensity reference. To ensure reproducibility, (i) high-quality graphite samples must be used such as HOPG grade ZYA or SCG, (ii) the D-band must be absent from the measured reference spectrum, (iii) optimized focus conditions must be used and (iv) accurate laser incident power and acquisition time normalizations must be performed. The experimental setup is fully enclosed to avoid any external perturbations. Together with its designed great mechanical and laser pointing stabilities, this allows to almost cancel any xyz drifts typically due to ambient temperature changes. A home-made data analysis software is used to treat the data set (including normalization of the intensity with regards to that of HOPG, corrections of the laser fluctuations, background subtraction, fitting of the bands, etc.).

**X-ray diffraction**

The crystallographic structure of the samples, fixed to silicon substrates to make them flat, has been analyzed by means of an X'Pert PRO Panalytical apparatus using the CuKα radiation (1.54056 Å).

**Hall bar fabrication**

Devices are fabricated after the CVD graphene is transferred onto a 7×7 mm$^2$ SiO$_2$ (300 nm)/p$^{++}$ Si substrate in a wet transfer procedure. A double layer resist mask (PMMA/MA and PMMA 950K) is fabricated using a customized electron beam lithography platform from Raith GmbH to define the contact structures, which are metallized with a titanium adhesion layer (5 nm) and gold (30 nm) contacts using a molecular beam epitaxy system at evaporation rates of ~ 0.1 nm/s and 0.02 nm/s, respectively. Afterwards, a second resist mask (single layer PMMA 950K) is employed as an etch mask in order to define the graphene transport channel using reactive ion etching (channel of width W = 10 μm and length L = 12 μm). Each of these two lithography steps is followed by lift-off in acetone.

**Transport measurements**

Electronic transport measurements are carried out in an Oxford Instruments He Heliox cryostat with a base temperature of 300 mK and a 5 T magnet. Current bias and gate voltage are applied using a Keithley model 2612 dual channel source meter, while transversal and longitudinal voltage are probed using a Hewlett-Packard 34420A nanovolt meter. The vacuum chamber containing sample is evacuated down to ~5×10$^{-6}$ mbar.